\documentclass[draftcls,12pt,onecolumn]{IEEEtran}

\usepackage{url,hyperref,subcaption}
\usepackage{textcomp}
\usepackage{xcolor}
\usepackage{amsmath}
\usepackage{amssymb}

 \usepackage[pdftex]{graphicx}
\graphicspath{{./images/}}

\newcommand{\Nu}{N_\text{u}}
\newcommand{\Nb}{N_\text{b}}

\begin{document}

\title{A hybrid beamforming design for massive MIMO LEO satellite communications} 
\author{Joan Palacios\,$^{1}$, Nuria Gonz\'{a}lez-Prelcic\,$^{1}$, Carlos Mosquera\,$^{2}$, Takayuki Shimizu\,$^{3}$,and Chang-Heng Wang\,$^{3}$\\
$^{1}$Electrical and Computing Engineering Department, North Carolina State University,  Raleigh, NC, USA \\
$^{2}$atlanTTic Research Center, Universidade de Vigo,  Vigo, Spain\\
$^{3}$ Toyota Motor North America, Mountain View, CA, USA}


\maketitle

\begin{abstract}
5G and future cellular networks intend to incorporate low earth orbit (LEO) satellite communication systems (SatCom) to solve the coverage and
availability problems that cannot be addressed by satellite-based or ground-based infrastructure alone. This integration of terrestrial and non terrestrial networks poses many technical challenges which need to be identified and addressed. To this aim, we design and simulate the downlink of a LEO SatCom compatible with 5G NR, with a special focus on the design of the beamforming codebook at the satellite side. The performance of this approach is evaluated for the link between a LEO satellite and a mobile terminal in the Ku band, assuming a realistic channel model and commercial antenna array designs, both at the satellite and the terminal. Simulation results provide insights on open research challenges related to analog codebook design and hybrid beamforming strategies, requirements of the antenna terminals to provide a given SNR, or required  beam reconfiguration capabilities among others.
\end{abstract}

\begin{IEEEkeywords} Beam codebook, massive MIMO, LEO satellite communications, non-terrestrial networks, hybrid beamforming.
\end{IEEEkeywords}

\section{Introduction}
Integrated satellite-terrestrial cellular networks are currently being pursued for long awaited commercial applications. The main expected benefit is an uninterrupted coverage, even in unserved/underserved areas. LEO constellations seem to be the most promising platforms for satellite-based non-terrestrial networks (NTNs), due to their relatively shorter propagation delay \cite{Kodheli2017}. Although this propagation delay between ground terminals and LEO satellites (3 ms at 1000 km above the Earth surface) is non-negligible, LEO constellations can support continuity and ubiquity of the radio services, and also latency critical applications with requirements within tens of ms. 

Beam configuration and reconfiguration in a link between a LEO satellite and a ground mobile terminal is challenging due to the long round trip delay, fast movement of the satellite and limited on board processing capabilities. Current LEO constellations \cite{delPortillo2019} make use of fixed analog beams that illuminate a given area of the Earth's surface, without any capability to steer narrow beams in the user directions. Digital precoding stages are designed independently of the analog beam with the purpose of reducing inter-beam interference among the beams illuminating adjacent regions of interest (ROI) \cite{Devillers2011}.  Moreover, these solutions assume that a single satellite is illuminating a large specific ROI, and handover procedures are put in place to enable satellite switching at the user side before the satellite movement causes loss of coverage. 

Designing the LEO satellite footprint and the specific beam codebook at the satellite side in an integrated satellite-terrestrial network is challenging. The potential beampatterns need to be designed to guarantee coverage of the Earth's surface, limit the inter-beam interference, maximize system throughput, and, at the same time, exhibit some degree of compatibility with conventional codebooks defined  in cellular standards. Commercial efforts on deploying LEO constellations, not necessarily for integration with the terrestrial network,  are described in \cite{delPortillo2019,Xia2019}. Sizes of the beam footprints for the different deployments are provided, but the details of the beam codebook are not available. The massive MIMO LEO satellite communication system designed in \cite{You2020} considers the set of all feasible beams generated by a uniform  planar array at the satellite rather than a codebook, without trying to reduce the inter-beam interference or the amount of on-board processing.

In this paper, we propose and evaluate a massive MIMO LEO satellite communication system operating in the Ku band, based on a hybrid beamforming architecture. In particular, we design first the footprint associated to a given LEO satellite. Then, we design the beam codebook for the hybrid beamforming stage such that the derived footprint is covered by the beams in the codebook. We  build our design using a 2-D DFT-based grid of beams as in Type I and Type II 5G New Radio CSI codebooks \cite{Miao2018,R1-1705926,R1-1708699,R1-1709232,TR38.211,TR38.214}., considering an oversampling factor to shape the available beampatterns to the size of the satellite  footprint.  Assuming this codebook at the satellite and commercially available SatCom antennas for the mobile terminals, we evaluate the coverage and throughput of this LEO massive MIMO system operating in the Ku band. Additional performance metrics such as SNR degradation over time or inter-beam interference are also evaluated. Numerical results  provide the baseline performance of a state-of-the art  massive MIMO LEO satellite communication system and allow the identification of open research challenges.

\section{System model}

\begin{figure}[ht]
    \centering
    \includegraphics[width = 0.8\linewidth]{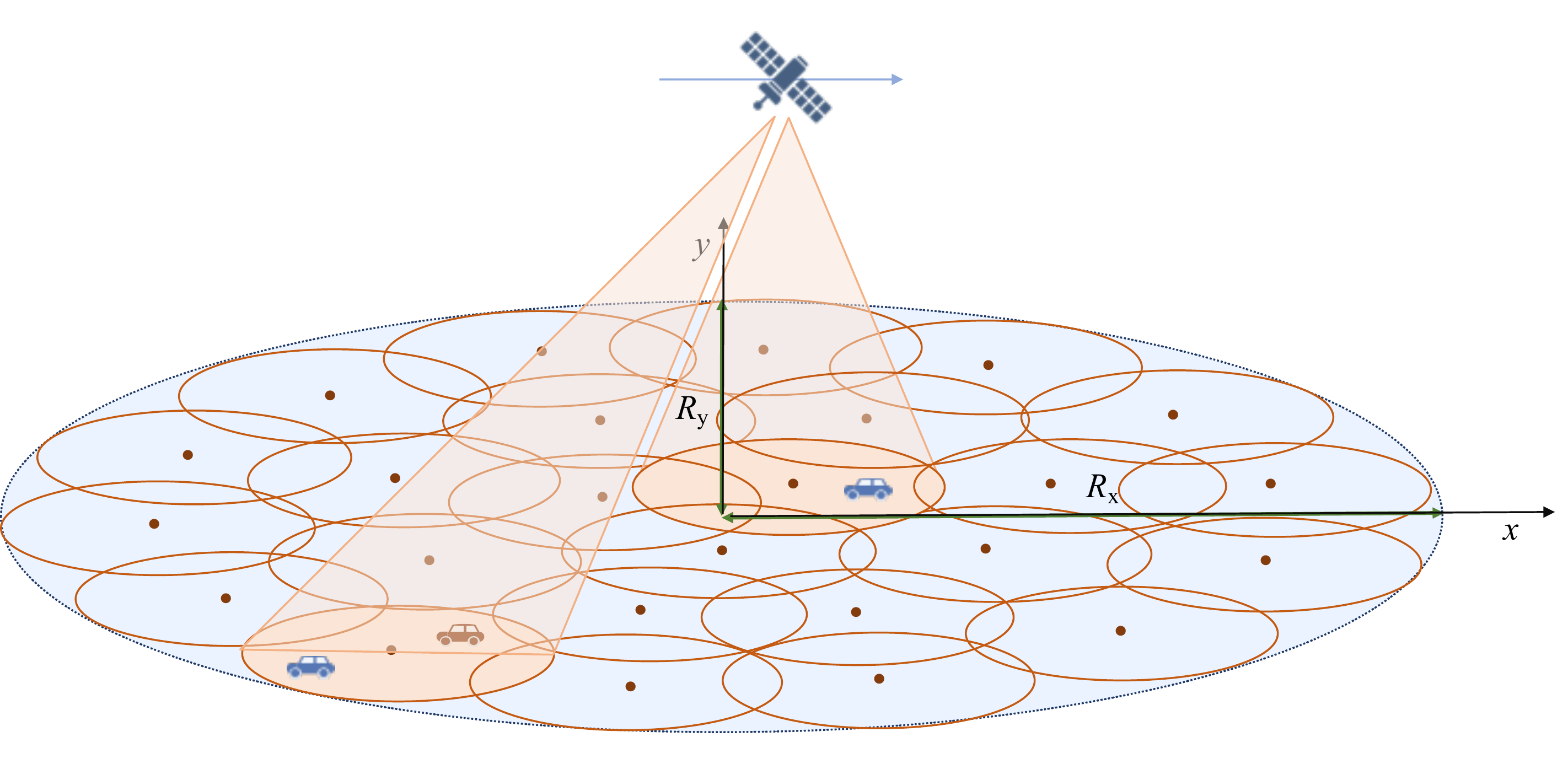}
    \caption{System model for a LEO satellite communication system covering an elliptical area using a predefined beam codebook.}
    \label{fig:system-model}
\end{figure}

We consider the downlink of a massive MIMO LEO satellite communication system operating in the Ku band, as in current commercial deployments \cite{delPortillo2019}, and also compliant with the 3GPP proposal for 5G broadband satellite services \cite{Angeletti2020}.  We denote as $B_{\rm DL}$ the  bandwidth allocated to the downlink. 
The system model we propose is general, but in the performance evaluation section we will focus on the {\em Communications on the move} (COOM) 5G use case. A given LEO satellite in the constellation can cover an elliptical region of interest (ROI), with semi-radius $R_{\rm x}$ and $R_{\rm y}$, using a set of potential beams, as illustrated in Fig.~\ref{fig:system-model}.
Thus, the system supports the direct transmission  to a maximum number of simultaneous mobile user terminals (UTs) on  the ground, denoted as $\Nu$, by using $\Nb$ analog beams, with $\Nu<\Nb$.  Time or frequency division multiplexing mechanisms will enable beam sharing  when more than one user per beam is simultaneously served. The beam footprints are not necessarily identical, to account for the different elevation angles across beams. We assume that the beams generated at the LEO satellite belong to a fixed codebook, i.e. the beams move with the satellite, as considered in Scenario C1 and D2 in  the 3GPP technical report that specifies solutions for 5G NR to support NTNs \cite{TR38.821}.
We also assume that the UTs know the position and trajectory  of the satellites in the constellation at all times, as proposed in \cite{Kim2020}.
This information is periodically updated and broadcasted by the LEO satellite operations center to the cellular network.  Knowledge of the planned trajectory allows the UTs to obtain an estimate of the satellite position even at locations where the terrestrial network is not available.
It is out of the scope of this paper to consider the impact of the position or trajectory errors in the system model. 

\subsection{MIMO architecture and antenna models}
\begin{figure}[ht]
    \centering
    \includegraphics[width = 0.4\linewidth, trim={6.8cm 4cm 10cm 1.7cm}, clip]{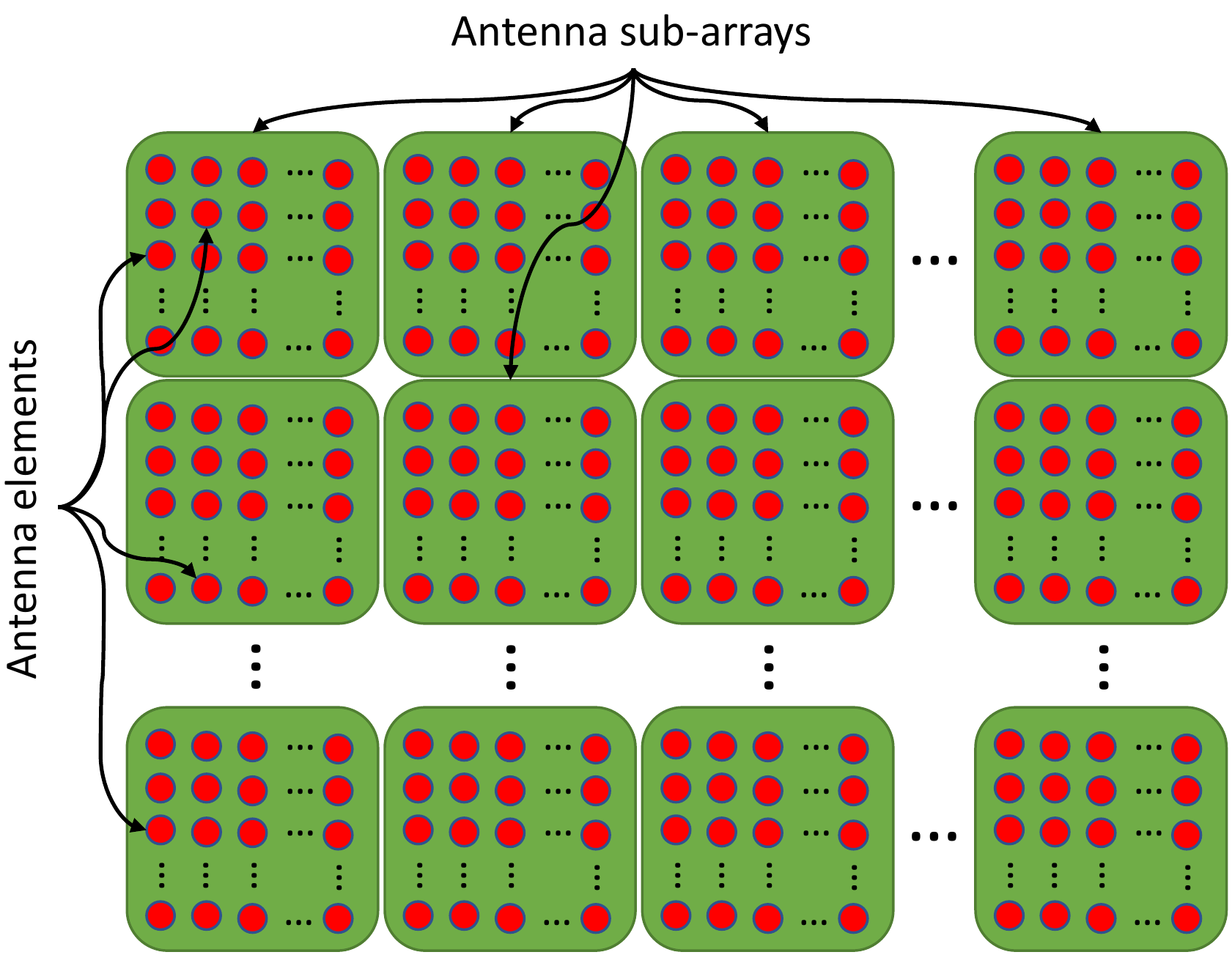}
    \caption{Partially connected sub-array structure at the satellite antenna. Each subarray is controlled by a single RF-chain.}
    \label{fig:SubArrays}
\end{figure}

We assume that the satellite is equipped with a large uniform planar array (UPA) with dimension $N^\text{sat}_{\rm x}\times N^\text{sat}_{\rm y}$ and $N^\text{sat} = N^\text{sat}_{\rm x}N^\text{sat}_{\rm y}$ denoting the total number of antennas. 
We consider ${\rm x}$ to be the axis in the direction of the satellite's movement and ${\rm y}$ to be the axis in the direction orthogonal to the movement.
The possible beamforming architectures for the LEO satellite include: fully digital, reconfigurable analog, analog based on beam switching  and hybrid \cite{Angeletti2020}.
Our choice is a partially connected hybrid MIMO architecture with a fixed codebook-based  analog stage. The number of beamforming ports in each dimension is defined as $N^\text{RF}_{\rm x}$ and $N^\text{RF}_{\rm y}$. The large array is organized into sub-arrays of size $N^\text{sub}_{\rm x}\times N^\text{sub}_{\rm y}$, with $N^\text{sub}_{\rm x} = N^\text{sat}_{\rm x}/N^\text{RF}_{\rm x},\quad N^\text{sub}_{\rm y} = N^\text{sat}_{\rm y}/N^\text{RF}_{\rm y}$ as illustrated in Fig.~\ref{fig:SubArrays}. Each radio frequency (RF) chain drives one of the subarrays, generating an independent spot beam so the whole coverage area is illuminated. This approach allows the reduction of the number of beamforming ports and on board processing requirements, since the analog beams do not have to be constantly reconfigured.   

Assuming a hybrid architecture, the precoding matrix ${\bf F}\in\mathbb{C}^{N_{\rm sat}\times \Nu}$ can be written as ${\bf F} = {\bf F}_{\rm RF}{\bf F}_{\rm BB}$, where ${\bf F}_{\rm BB}\in\mathbb{C}^{\Nb \times \Nu}$ is the digital precoder and  ${\bf F}_{\rm RF}\in\mathbb{C}^{N^\text{sat} \times \Nb}$
is the analog precoder. 
Due to power constraints in the $m$-th RF-chain, we have that $\|[{\bf F}_{\rm BB}]_{m, :}\|^2 \leq P_\text{RF}$. We define the index function $I_{\rm RF}: \{1, \ldots, N_{\rm sat}\}\rightarrow\{1, \ldots, N_{\rm b}\}$, where $I_{\rm RF}(n)$ is the index of the RF-chain to which the $n$-th antenna is connected. With this definition,  the RF precoder ${\bf F}_{\rm RF}$ can be written as $[{\bf F}_{\rm RF}]_{n, m} = \left\{\begin{array}{rl}
    e^{\phi_ni} & {\rm if } \quad m = I_{\rm RF}(n) \\
    0 & {\rm otherwise}
\end{array}\right.$.
We envision a digital precoder at the satellite to manage inter-beam interference in the context of a full frequency reuse system.   Finally, it is also assumed that all the power amplifiers at the satellite UPA are identical, and their impact can be modeled as a scaling factor $\sqrt{P}$ \cite{Angeletti2020}. 

For the antenna arrays at the mobile terminal, we consider three possible solutions that enable beamforming in the satellite direction.
The first one is a uniform planar array (UPA) of moderate size, with dimensions $N^\text{UT}_{\rm x}\times N^\text{UT}_{\rm y}$, having a total of $N^{\rm UT}=N^\text{UT}_{\rm x} N^\text{UT}_{\rm y}$ antenna elements.
Recent advances on antenna technology enable the commercial applications of these planar arrays for SatCom terminals \cite{SatixFy}. The second considered solution is a flat panel antenna based on leaky wave technology \cite{Mehdipour2014}. The final option is a commercial liquid crystal based metasurface antenna developed by Kymeta \cite{Mehdipour2019}.

We define ${\bf w}\in\mathbb{C}^{N^{\rm UT}}$ as the vector containing the UT antenna weights over the set of feasible configurations.
We consider first the UPA  with antenna elements relative positions defined through ${\bf k}_{\rm u}\in\mathbb{C}^{N_{\rm u}\times 3}$, being  $[{\bf k}_{\rm u}]_{n, :}$ the $n$-th antenna element's relative position. In this case, the UT steering vector ${\bf a}_{\rm u}:\mathcal{S}_2\rightarrow\mathbb{C}^{N_{\rm u}}$ is defined over the unitary sphere $\mathcal{S}_2$ as ${[{\bf a}_{\rm u}({\bf v})]_n} = e^{-i\frac{2\pi}{\lambda}<{[\bf k}_{\rm u}]_{n, :}, {\bf v}>}$. The maximum gain in a direction ${\bf v}\in\mathcal{S}_2$ can be computed as $\max_{\bf w}\frac{|<{\bf a}_{\rm u}({\bf v}), {\bf w}>|}{\|{\bf w}\|}$.  The set of feasible beam-patterns for a uniform planar array is $\{z\in\mathbb{C}: |z|=1\}^{N_{\rm u}}$, and its maximum achievable gain in any direction is $10\log_{10}(N^\text{UT})$ when each one of the antenna elements is assumed to provide a unitary gain.

For a leaky wave antenna with leakage factor $\alpha$,  a feasible beam-patterns ${\bf w}\in\mathbb{C}^{N^{\rm UT}}$ follows the expression 
$[{\bf w}]_n = e^{-\frac{1}{\lambda}([{\bf k}_{\rm u}]_{n, 1}\alpha + [{\bf k}_{\rm u}]_{n, 2}\alpha)-\frac{i}{\lambda}([{\bf k}_{\rm u}]_{n, 1}\theta_1 + [{\bf k}_{\rm u}]_{n, 2}\theta_2)}$ for $\theta_1, \theta_2\in[-\pi, \pi]$ \cite{Mehdipour2014}.
When $\alpha \neq 0$,  the gain of the leaky wave antenna can be computed as $10\log_{10}(\frac{(1-e^{-\alpha N^\text{UT}_{\rm x}})(1-e^{-\alpha N^\text{UT}_{\rm y}})(1+e^{-\alpha})(1+e^{-\alpha})}{(1+e^{-\alpha N^\text{UT}_{\rm x}})(1+e^{-\alpha N^\text{UT}_{\rm y}})(1-e^{-\alpha})(1-e^{-\alpha})})$.
Note that there is a gain threshold $10\log_{10}(\frac{(1+e^{-\alpha})(1+e^{-\alpha})}{(1-e^{-\alpha})(1-e^{-\alpha})})$ that cannot be exceeded by increasing the number of antenna elements in the leaky-wave antenna. 

Finally, the metasurface antenna designed by Kymeta  has a peak gain of $33$ dB and a $1.2$ cosine roll-off factor. This translates into a gain of $33-1.2\cdot10\log_{10}(\sin(\theta_{\rm el}))$ dB towards a satellite with an elevation angle $\theta_{\rm el}$ relative to the user. The analytical expression of the array factor for this antenna is not available, so only system metrics that depend on the peak gain can be computed when the terminal is operating with this antenna.

\subsection{Channel model}

We consider the channel between a single LEO satellite located at position ${\bf x}_{\rm sat}\in\mathbb{R}^3$ and a UT located at position ${\bf x}_{\rm u}\in\mathbb{R}^3$ on the Earth's surface, both measured from Earth's center.
The satellite is orbiting Earth at a height $h_{\rm sat}$,  moving with a linear speed of
\begin{equation}
v_{\rm sat} = \sqrt{G\frac{m_{\rm earth}}{R_{\rm earth}+h_{\rm sat}}},
\end{equation}
where $R_{\rm earth}$ is the Earth's radius, $m_{\rm earth}$ is the Earth's mass, and $G$ is the gravitational constant.
This is equivalent to an angular speed of $w_{\rm sat} = \frac{v_{\rm sat}}{r_{\rm earth}+h_{\rm sat}}$.
The closest point to the satellite on the Earth surface, from now on the satellite's shadow, moves at a linear speed of $v_{\rm shadow} = w_{\rm sat}r_{\rm earth}$ and circles the Earth with a period of $\frac{2\pi}{w_{\rm sat}}$.

\subsubsection{Link budget model}
The link power loss includes the free space path loss, denoted as $LP_{\rm fs}$, and the atmospheric absorption loss $LP_{\rm at}$.  We neglect the impact of the atmospheric fading on the link budget.
This way, the signal-to-noise ratio (SNR) can be written in terms of the the received signal strength (RSS) and the noise power $\sigma^2$ as \cite{TR38.821} 
\begin{align}
SNR{\rm [dB]}=RSS{\rm [dBw]}-\sigma^2{\rm [dBw]}\\
RSS{\rm [dBw]}=P_{\rm TX}{\rm [dBw]}-LP_{\rm cable}{\rm [dB]}+G_{\rm TX}{\rm [dB]}-LP_{\rm at}{\rm [dB]}-LP_{\rm fs}{\rm [dB]}+G_{\rm RX}{\rm [dB]}\\
\sigma^2{\rm [dBw]}=T{\rm [dBK]}+k{\rm [dBw/K/Hz]}+B{\rm [dBHz]},
\end{align}
where $P_{\rm TX}$ is the transmit power, $G_{\rm TX}$ is the transmit antenna gain, $LP_{\rm cable}$ is the cable loss between the antenna and the transmitter, $G_{\rm RX}$ is the receiver antenna gain, ${T}$ is the noise temperature, $k$ is the Boltzmann constant equal to $-228.6 {\rm dBW/K/Hz}$, and $B$ is the channel bandwidth.

The free space path loss in dB is given by
\begin{equation}\label{eq:path_loss}
LP_{fs} = 20(\log_{10}(\|{\bf x}_{\rm sat}-{\bf x}_{\rm u}\|)+\log_{10}(f)+\log_{10}(\frac{4\pi}{c})),
\end{equation}
with $\|{\bf x}_{\rm sat}-{\bf x}_{\rm u}\|$ being the traveled distance, $f$ the frequency and $c$ the speed of light.

\begin{figure}
    \centering
    \includegraphics[width = 0.4\linewidth, trim={4cm 8.5cm 4cm 9cm}, clip]{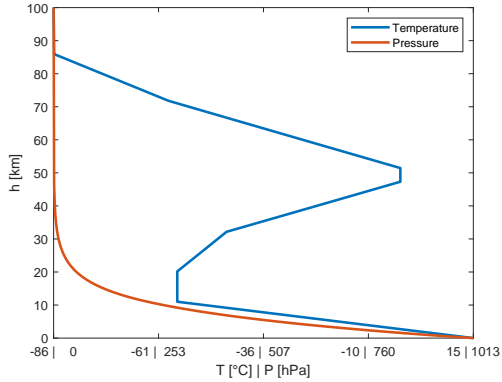}
    \caption{1976 Standard Atmospheric model showing temperature and pressure vertical profiles.}
    \label{fig:atmos}
\end{figure}

Defining $A(f, T(h), P(h), \rho(h))$ as the loss per meter for temperature $T(h)$, pressure $P(h)$ and humidity $\rho(h)$ at height $h$, the  atmospheric path loss due to absorption  is
\begin{equation}
LP_{at} = \int_{h_{\rm min}}^{h_{\rm max}} A(f, T(h), P(h), \rho(h))dh.
\end{equation}
The integral is performed over the efficient limits of the atmosphere, that is between the user's height $h_{\rm min}$ ($h_{\rm min} = 0$ for sea level) and the atmosphere thickness $h_{\rm max} = 81km$.
The atmospheric temperature and pressure as a function of the height can be extracted using the 1976 Standard Atmospheric model depicted in Fig.~\ref{fig:atmos}, while the corresponding loss per meter can be computed using tabulated functions implemented in available software packages like {\tt gaspl} in MATLAB.
The humidity density vertical profile $\rho(h)$ can significantly vary depending on meteorological conditions, and its physical properties (condensation and freezing) can barely exist past its freezing point ($< -0$ \textdegree{}C) without the formation of clouds. Therefore, we will assume that there are no clouds and that the water vapor is $7.5 {\rm g}/{\rm m}^3$ at sea level $h=0$ and proportional to the air density in the first $2.3$ km of the atmosphere, where the temperature is still over the $0\deg$.

\subsubsection{Satellite angular speed and Doppler effects} \label{sec:Doppler_angular}
The satellite movement at a speed ${\bf v}_{\rm sat}\in\mathbb{R}^3$ introduces a relative angular speed $w_{\rm rel}$ with respect to the user and, for a signal at frequency $f$, a Doppler effect $\Delta f_{\rm sat}$ with expressions
\begin{equation}\label{eq:doppler_ang}
\begin{array}{rl}
    \Delta f_{\rm sat} = & <{\bf v}_{\rm sat}, \frac{{\bf x}_{\rm u}-{\bf x}_{\rm sat}}{\|{\bf x}_{\rm u}-{\bf x}_{\rm sat}\|}>\frac{f}{c}, \\
    w_{\rm rel} = & \sqrt{1-|<\frac{{\bf v}_{\rm sat}}{v_{\rm sat}}, \frac{{\bf x}_{\rm u}-{\bf x}_{\rm sat}}{\|{\bf x}_{\rm u}-{\bf x}_{\rm sat}\|}>|^2}\frac{v_{\rm sat}}{\|{\bf x}_{\rm u}-{\bf x}_{\rm sat}\|}.
\end{array}
\end{equation}
Note that the angular speed is computed in absolute terms and it is not directly related to a parametrization of choice. like for example polar coordinates. Therefore, it does not translate to variational speed of azimuth and elevation angles.
We discourage the use of satellite tracking strategies in azimuth and elevation to avoid sharp changes in the azimuth direction. For example, 
a small change of direction  from $[\epsilon, 0, \sqrt{1-\epsilon^2}]$ to $[-\epsilon, 0, \sqrt{1-\epsilon^2}]$ for any $\epsilon\in[0, 1]$ as small as desired in the real world close to the point of maximum elevation, involves a $180^\circ$ change in the azimuth angle, while the effective change of direction is only  
$\arccos(1-2\epsilon^2) {\rm [rad]}$, that for small values of $\epsilon$ is close to  $2\epsilon {\rm [rad]}$.

The Doppler effect caused by the satellite's movement is increasing and anti-symmetric in the movement's direction, thus its maximum is located at the elliptical footprint boundary.
By creating orbits with many satellites, we can reduce the maximum absolute Doppler effect, since the distance between satellites decreases, thus reducing the size of the covered area in the direction of movement.
As for the relative angular speed, the effect is symmetric concave with a maximum in the satellite's shadow.
Therefore, the maximum is not affected by the satellite's density.
Using basic geometry, these maximum values are given by the following formulas
\begin{equation}\label{eq:doppler_ang_max}
\begin{array}{rl}
    \max |\Delta f_{\rm sat}| = & \frac{\frac{R_{\rm earth}R_{\rm x}}{\sqrt{R_{\rm earth}^2+R_{\rm x}^2}}\sqrt{\frac{Gm_{\rm earth}}{R_{\rm earth}+h_{\rm sat}}}}{\sqrt{R_{\rm earth}^2+(R_{\rm earth}+h_{\rm sat})^2-2\frac{R_{\rm earth}^2(R_{\rm earth}+h_{\rm sat})}{\sqrt{R_{\rm earth}^2+R_{\rm x}^2}}}}\frac{f}{c}, \\
    \max w_{\rm rel} = & \frac{1}{h_{\rm sat}}\sqrt{G\frac{m_{\rm earth}}{R_{\rm earth}+h_{\rm sat}}},
\end{array}
\end{equation}
The maximum Doppler caused by the satellite's movement can be simplified when $R_{\rm x} << R_{\rm earth}$ to
\begin{equation}
\max |\Delta f_{\rm sat}| = \frac{R_{\rm x}}{h_{\rm sat}}\sqrt{\frac{Gm_{\rm earth}}{R_{\rm earth}+h_{\rm sat}}}\frac{f}{c}= \frac{R_{\rm x}v_{\rm sat}}{h_{\rm sat}}\frac{f}{c}.
\end{equation}

The Doppler effect caused by the UT moving with the speed vector ${\bf v}_{\rm u}\in\mathbb{R}^3$, for a random ${\bf v}\in\mathcal{S}_2$ which we are assuming to be uniformly distributed, is given by $\Delta f_{\rm u} = <{\bf v}_{\rm u}, {\bf v}>\frac{f}{c}$.
The maximum average value for this expression is given by $\max|\Delta f_{\rm u}| = v_{\rm u}\frac{f}{c}$, while the average absolute value can be computed as $\mathbb{E}(|\Delta f_{\rm u}|) = \frac{v_{\rm u}}{6}\frac{f}{c}$, being $v_{\rm u} = \|{\bf v}_{\rm u}\|$ the absolute UT speed.

\subsubsection{Channel matrix}
Due to the large distance and lack of medium interaction, a Rician  geometric channel model with power $\beta$ and Rician factor $K_{\rm r}$ can be assumed \cite{You2020}. We will also assume that the Doppler effect can be corrected and removed  from the channel matrix. The UPA at the satellite can be described by steering vectors in the direction ${\bf v}\in\mathcal{S}_2$, denoted as ${\bf a}_{\rm sat}: \mathcal{S}_2\rightarrow\mathbb{C}^{N^{\rm sat}}$. The antenna element positions are defined as ${\bf k}_{\rm sat}\in\mathbb{R}^{N^{\rm sat}\times3}$ with $[{\bf k}_{\rm sat}]_{n, :}$ being the satellite's $n$-th antenna element relative position. With this definition,  the steering vectors can be written as ${[{\bf a}_{\rm sat}({\bf v})]_n} = e^{-i\frac{2\pi}{\lambda}<{[\bf k}_{\rm sat}]_{n, :}, {\bf v}>}$.
An alternative notation considering the spherical coordinates for the azimuth ($\phi$) and elevation ($\theta$) angles can be obtained by 
substituting ${\bf v} = [\cos(\phi)\cos(\theta), \sin(\phi)\cos(\theta), \sin(\theta)]$. This way, the steering vectors are written as ${[{\bf a}_{\rm sat}(\phi, \theta)]_n} = e^{-i\frac{2\pi}{\lambda}({[\bf k}_{\rm sat}]_{n, 1}\cos(\phi)\cos(\theta)+[{\bf k}_{\rm sat}]_{n, 2}\sin(\phi)\cos(\theta)}$.
Given the steering vectors, the equivalent channel matrix after Doppler correction is given by
\begin{equation}\label{eq:rician_channel}
{\bf H} = (\sqrt{\frac{K_{\rm r}\beta}{K_{\rm r}+1}}{\bf a}_{\rm u}(\frac{{\bf x}_{\rm x}-{\bf x}_{\rm sat}}{\|{\bf x}_{\rm u}-{\bf x}_{\rm sat}\|})+\sqrt{\frac{\beta}{K_{\rm r}+1}}{\bf a}_{\rm R}){\bf a}_{\rm sat}^{\rm H}(\frac{{\bf x}_{\rm sat}-{\bf x}_{\rm u}}{\|{\bf x}_{\rm sat}-{\bf x}_{\rm u}\|}),
\end{equation}
where ${\bf a}_{\rm R}\sim\mathcal{CN}(0, \Sigma)$ is the Rician component satisfying ${\rm trace}(\Sigma) = \|{\bf a}_{\rm u}(\frac{{\bf x}_{\rm u}-{\bf x}_{\rm sat}}{\|{\bf x}_{\rm u}-{\bf x}_{\rm sat}\|})\|^2$. Defining $\gamma=\sqrt{\frac{K_{\rm r}\beta}{K_{\rm r}+1}}$ as the complex gain comprising all path loss effects and phase shift corresponding to the LoS component, the channel matrix can be written in a more compact way as 
\begin{equation}\label{eq:channel}
{\bf H} = \gamma({\bf a}_{\rm u}(\frac{{\bf x}_{\rm u}-{\bf x}_{\rm sat}}{\|{\bf x}_{\rm u}-{\bf x}_{\rm sat}\|})+\frac{1}{\sqrt{K_{\rm r}}}{\bf a}_{\rm R}){\bf a}_{\rm sat}^{\rm H}(\frac{{\bf x}_{\rm sat}-{\bf x}_{\rm u}}{\|{\bf x}_{\rm sat}-{\bf x}_{\rm u}\|}).
\end{equation}

\section{System design}

\subsection{Design of the coverage area}
\begin{figure}
    \centering
    \begin{tabular}{cc}
    \includegraphics[width = 0.3\linewidth, trim={5cm 7.5cm 5cm 6.5cm}, clip]{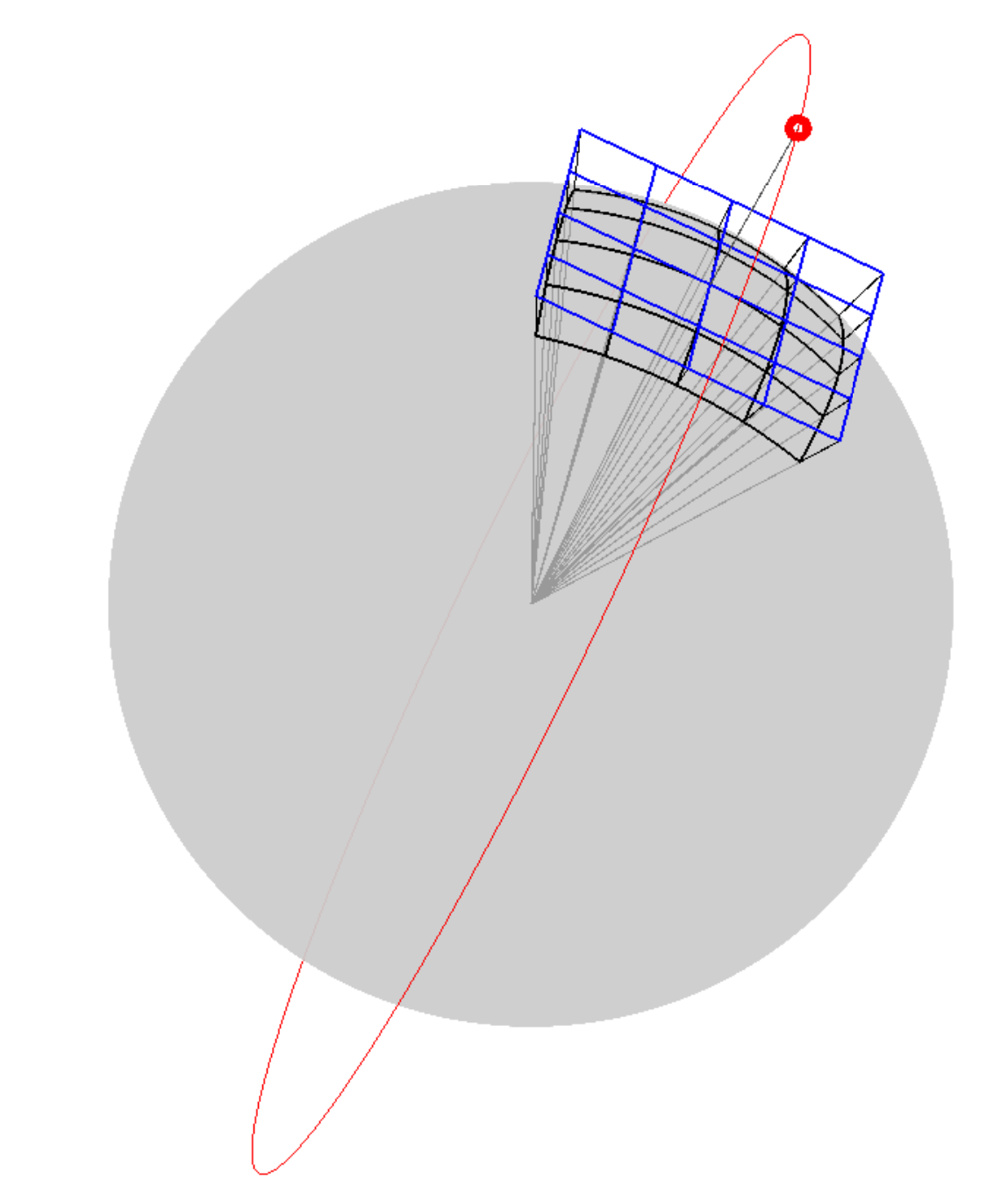} &  \includegraphics[width = 0.4\linewidth]{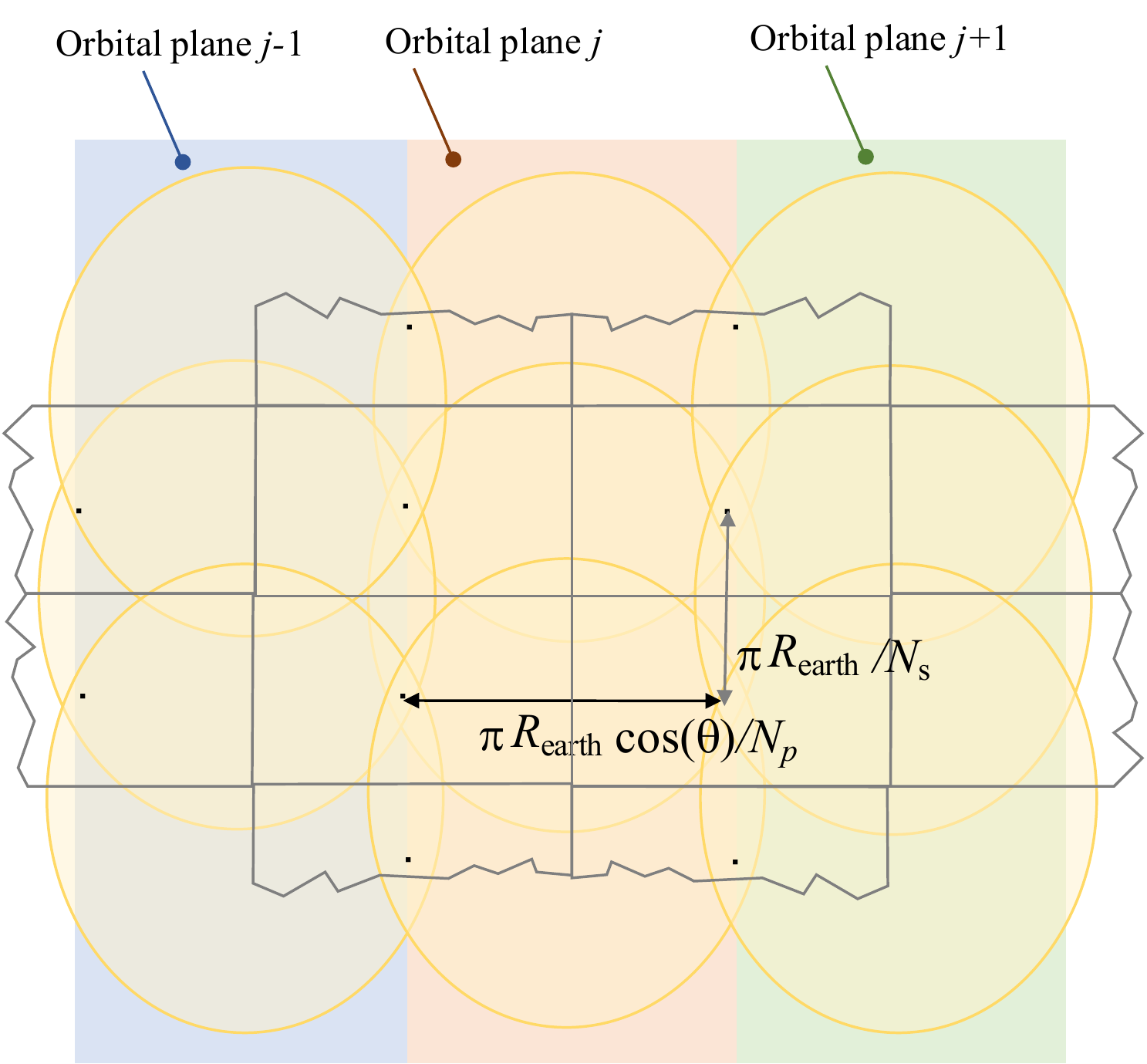} \\
    (a) & (b) \\
    \end{tabular}
    \caption{Design of the satellite elliptical footprint. (a) Stereographic parametrization of the area covered by the satellite. (b) Illustration of the design process.}
    \label{fig:stereo}
\end{figure}

We consider a LEO constellation where satellites are distributed around $N_{\rm p}$ orbital planes with an inclination $\theta_{\rm op}$, each one with $N_{\rm s}$ satellites.
Each satellite illuminates an elliptical area as shown in Fig.~\ref{fig:system-model}. 
From a geometric perspective, to parametrize the area covered by the satellite, we are choosing a stereographic projection respect to the Earth's center as illustrated in Fig.~\ref{fig:stereo}(a).
This allows us to use Cartesian coordinates while still considering Earth's intrinsic geometry.
Mathematically, this parametrization of the covered area translates the Cartesian point $(x, y)$ into the sphere point
$R_{\rm earth}\frac{R_{\rm earth}{\bf u}_{\rm sat} + x{\bf u}_{\rm x} + y{\bf u}_{\rm y}}{\sqrt{R_{\rm earth}^2+x^2+y^2}}$ for the unitary vector definition ${\bf u}_{\rm sat} = \frac{{\bf x}_{\rm sat}}{\|{\bf x}_{\rm sat}\|}$, ${\bf u}_{\rm x} = \frac{{\bf v}_{\rm sat}}{\|{\bf v}_{\rm sat}\|}$ and ${\bf u}_{\rm y} = {\bf u}_{\rm x} \wedge {\bf u}_{\rm sat}$ using $\wedge$ to denote the cross product.

To design the sizes of the semiradiuses of the elliptical ROI,  we will force the LEO constellation to cover the entire Earth. 
Thus, the equivalent flat problem we want to solve is to cover the entire Earth surface with ellipses centered in the vertices of a rectangular grid with spacing $\pi R_{\rm earth}/N_{\rm s}$  in the first dimension and $\pi R_{\rm earth}\sin(\theta_{\rm op})/N_{\rm p}$  in the second dimension. Due to ellipses being convex and symmetric in both axes, this is equivalent to design ellipses containing the center points of the rectangular grid, as illustrated in Fig.~\ref{fig:stereo}(b). 
This condition can be written as 
\begin{equation}
\left(\frac{\pi R_{\rm earth}/N_{\rm s}}{R_{\rm x}}\right)^2+\left(\frac{\pi R_{\rm earth}\sin(\theta_{\rm op})/N_{\rm p}}{R_{\rm y}}\right)^2 \leq 1.
\end{equation}
An example of a semi-radius pair that fulfills  this condition is
\begin{equation}
R_{\rm x} = \sqrt{2}\pi R_{\rm earth}/N_{\rm s}, \,\,\, R_{\rm y} = \sqrt{2}\pi R_{\rm earth}\sin(\theta_{\rm op})/N_{\rm p}.
\end{equation}

\subsection{Beam codebook design}\label{sec:codebook}

\begin{figure}[ht]
\centering
\includegraphics[width = 0.5\linewidth]{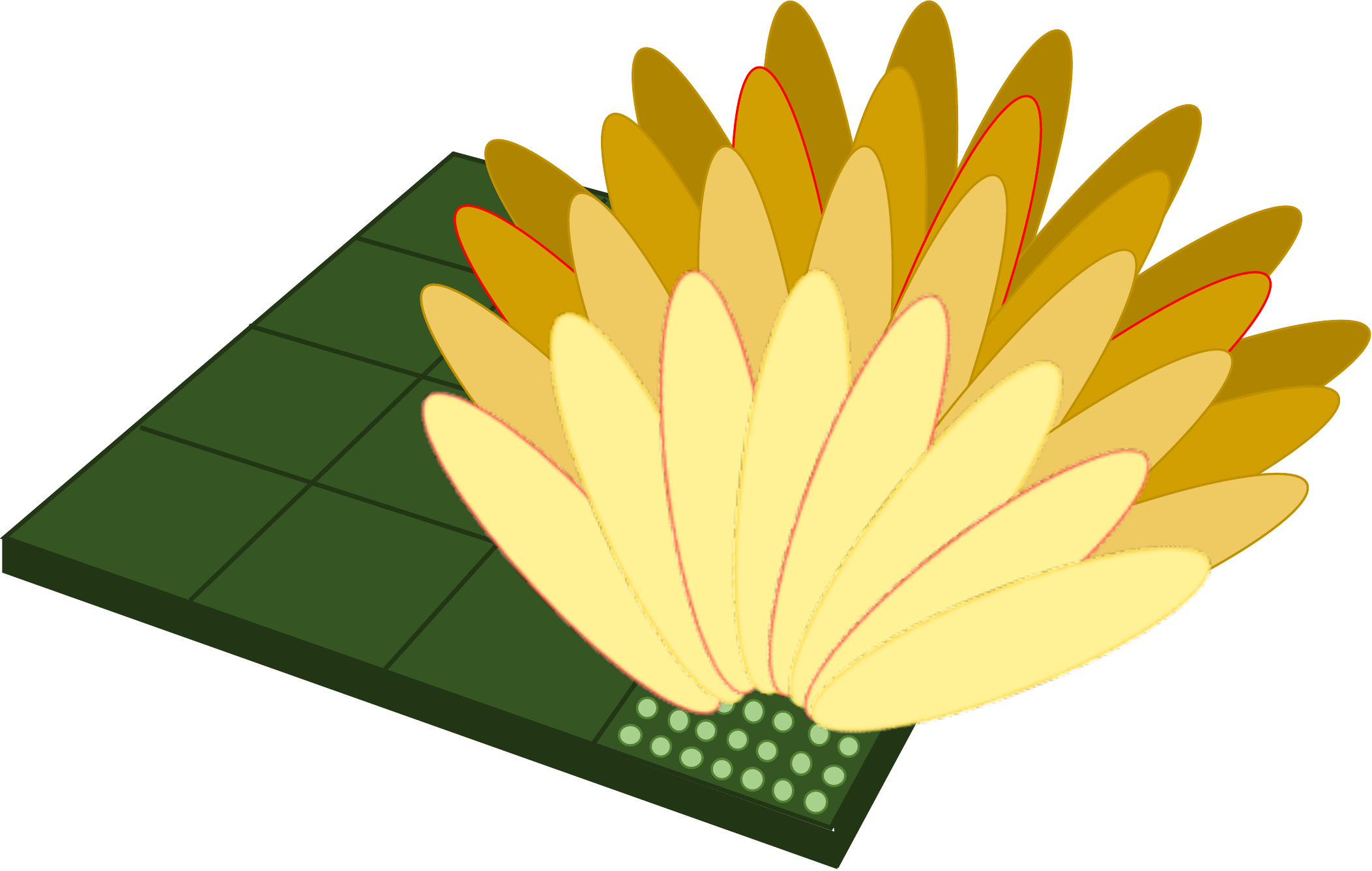}
\caption{Illustration of the GoB generated with a 2D DFT codebook  with an oversampling factor of 2. Orthogonal beams are marked in red.}
\label{fig:GoB}
\end{figure}

Our goal is to design a set of analog beampatterns or beam codebook, such that the set of beam footprints covers an ellipsoidal ROI as that described in the previous section. Depending on the location of the active users, some active beams will be selected from the set available in the codebook.  We will consider a DFT-based analog precoder codebook structure, a popular solution when operating with planar arrays. In this case, the array response vectors in each dimension have a linearly increasing phase, like the columns in the DFT matrix.  Therefore, the MIMO channel can be written in terms of array response vectors with this structure. DFT-based precoders match in this case the channel structure providing a good beamforming gain. Note that the column vectors in a DFT codebook generate a grid of beams (GoB) that can be used to span all angular directions, as illustrated in Fig.~\ref{fig:GoB} for a planar array with 16 panels (subarrays), and $7\times 5$ antenna elements per subarray. 

In particular, 5G New Radio defines what is called a Type II DFT codebook, where an oversampling factor is introduced in the definition of array steering vectors for azimuth and elevation \cite{Miao2018,R1-1705926,R1-1708699,R1-1709232,TR38.211,TR38.214}. The goal of the oversampling factor, which is fixed and equal to 4 in 5G NR, is to mitigate the loss of beamforming gain in directions between orthogonal beams (straddling loss). This way, beam orthogonality is lost in general, and only beams spaced the oversampling factor are orthogonal. Fig.~\ref{fig:GoB} shows the set of orthogonal beams marked in red when an oversampling factor equal to 2 has been used in each dimension.

In our design, we will also consider a 2D oversampled DFT codebook for each subarray, although the oversampling factor $O$ is not fixed, allowing the adjustment of the number of beams that illuminate footprints of different sizes for some given RF resources.
The elements of the block diagonal RF precoding matrix ${\bf F}_{\rm RF}\in\mathbb{C}^{N^{\rm sat}\times N_{\rm b}}$ can be written in terms of this oversampled DFT codebook as
\begin{equation}
[{\bf F}_{\rm RF}]_{n, I_{\rm RF}(n)} = e^{-i\frac{2}{\lambda}([{\bf k}_{\rm sat}]_{n, 1}\theta_{I_{\rm RF}(n), 1}+[{\bf k}_{\rm sat}]_{n, 2}\theta_{I_{\rm RF}(n), 2})},
\end{equation}
for $\{(\theta_{m, 1}, \theta_{m, 2})\}_m \subseteq \{\frac{2\pi(n-1)}{ON_{\rm x}^{\rm sub}}\}_n^{ON_{\rm x}^{\rm sub}}\times\{\frac{2\pi(n-1)}{ON_{\rm y}^{\rm sub}}\}_n^{ON_{\rm y}^{\rm sub}}$. Note that the $m$-th element of the codebook is contained in the non-zero entries of the $m$-th column of ${\bf F}_{\rm RF}$, that is $[{\bf F}_{\rm RF}]_{n, m} = e^{-i\frac{2}{\lambda}([{\bf k}_{\rm sat}]_{n, 1}\theta_{m, 1}+[{\bf k}_{\rm sat}]_{n, 2}\theta_{m, 2})}$.
This expression is also proportional to the steering vector in the direction ${\bf v}\in\mathcal{S}_2$ with $[{\bf v}]_1 = \frac{\theta_{m, 1}}{\pi}$ and $[{\bf v}]_2 = \frac{\theta_{m, 2}}{\pi}$.
To create the codebook, we prune the complete oversampled DFT dictionary, keeping only those beam-patterns  pointing to the elliptical ROI. The oversampling factor can be used to adapt the RF resources to the size of the ROI and is the key parameter that drives the tesselation of the coverage area. By increasing $O$, the beam footprint is reduced, and the number of beampatterns pointing to the desired ROI increases. The optimal value for $O$, in terms of provided SNR, allows the creation of as many beams pointing to the ROI as available RF chains. Note that when increasing the oversampling factor, the dictionary provides a higher SNR over the ROI because the straddling loss decreases,  but the interference between adjacent beams increases, leading to a reduction of the SINR. Therefore, the oversampling factor has to be carefully selected to find the appropriate tradeoff between SNR and SINR. The inter-beam interference problem can be addressed by designing  a suitable digital precoding matrix ${\bf F}_{\rm BB}$ that enables full frequency reuse, or by allocating different carriers to users served by adjacent beams, as proposed in \cite{Angeletti2020}. 

Once the codebook is created, the cells are defined as the areas inside the elliptical ROI where a given beam provides a higher gain than any other beam in the available codebooks for all the subarrays. Note that the DFT nature of the designed codebook creates a rectangular tiling of the ROI.
Section~\ref{sec:simulations} includes examples of the beam footprints when this type of codebook is assumed on board.

\section{Coverage  and spectral efficiency calculation}

\subsection{Coverage}\label{sec:coverage}
When operating with a transmitting beam-pattern ${\bf f}\in\mathbb{C}^{N^{\rm sat}}$ in the satellite beam codebook, and a receiving beam-pattern ${\bf w}\in\mathbb{C}^{N^{\rm UT}}$, the RSS is given by
\begin{equation}
RSS{\rm [W]} = |{\bf w}^{\rm H}{\bf H}{\bf f}|^2 = |\gamma|^2|<{\bf a}_{\rm u}(\frac{{\bf x}_{\rm u}-{\bf x}_{\rm sat}}{\|{\bf x}_{\rm u}-{\bf x}_{\rm sat}\|})+\frac{1}{\sqrt{K_{\rm r}}}{\bf a}_{\rm R}, {\bf w}>|^2|<{\bf a}_{\rm sat}^{\rm H}(\frac{{\bf x}_{\rm sat}-{\bf x}_{\rm u}}{\|{\bf x}_{\rm sat}-{\bf x}_{\rm u}\|}), {\bf f}>|^2.
\label{eq:RSS}
\end{equation}
Note that this expression depends on the user and satellite positions in  ${\bf x}_{\rm u}$ and ${\bf x}_{\rm sat}$ respectively.
The different terms in (\ref{eq:RSS}) can be identified with the terms in the link budget equation. Thus  $G_{\rm RX}{\rm [dB]} = 10\log_{10}(|<{\bf a}_{\rm u}(\frac{{\bf x}_{\rm u}-{\bf x}_{\rm sat}}{\|{\bf x}_{\rm u}-{\bf x}_{\rm sat}\|})+\frac{1}{\sqrt{K_{\rm r}}}{\bf a}_{\rm R}, {\bf w}>|^2)$, the transmitter gain $G_{\rm TX}{\rm [dB]} = 10\log_{10}(|<{\bf a}_{\rm sat}^{\rm H}(\frac{{\bf x}_{\rm sat}-{\bf x}_{\rm u}}{\|{\bf x}_{\rm sat}-{\bf x}_{\rm u}\|}), {\bf f}>|^2)$, and $10\log_{10}(|\gamma|^2)$ comprises all the losses $LP_{\rm cable}[dB]+LP_{\rm at}[dB]+LP_{\rm fs}[dB]$.

To compute the combining gain at the UT side, we will consider the LoS direction to be known (in the system model we have assumed that the satellite position is known) and the Rician component of the channel to be unknown.  Thus, the combiner ${\bf w}_\text{u}$ will be computed as the one maximizing $\frac{|<{\bf a}_{\rm u}({\bf v}), {\bf w}>|^2}{\|{\bf w}\|^2}$ for the set of feasible ${\bf w}$ as discussed in Section 2.1.
The average gain associated to this combiner can be computed as $\mathbb{E}(\frac{|<{\bf a}_{\rm u}({\bf v})+\frac{1}{\sqrt{K_{\rm r}}}\bf a_{\rm R}, {\bf w}_{\rm u}>|^2}{\|{\bf w}_{\rm u}\|^2})$.
Since ${\bf a}_{\rm R}$ is Gaussian and independent of any other variable, we can write the average gain as
\begin{equation}
\mathbb{E}(\frac{|<{\bf a}_{\rm u}({\bf v})+\frac{1}{\sqrt{K_{\rm r}}}\bf a_{\rm R}, {\bf w}_{\rm u}>|^2}{\|{\bf w}_{\rm u}\|^2})=\mathbb{E}(\frac{|<{\bf a}_{\rm u}({\bf v}), {\bf w}_{\rm u}>|^2}{\|{\bf w}_{\rm u}\|^2})+\frac{1}{K_{\rm r}}\mathbb{E}(\frac{|<{\bf a}_{\rm R}, {\bf w}_{\rm u}>|^2}{\|{\bf w}_{\rm u}\|^2}).
\end{equation}
Now, if we make use of the definition of ${\bf w}_{\rm u}$ and consider the Rician component to be uniform, i.e.  $\Sigma = {\bf I}$, we obtain the final expression for the receive gain as
\begin{equation}
\mathbb{E}(\max_{{\bf w}}\frac{|<{\bf a}_{\rm u}({\bf v}), {\bf w}>|^2}{\|{\bf w}\|^2})+\frac{1}{K_{\rm r}}.
\end{equation}

Since the LEO satellite is transmitting simultaneously through multiple beams with a gain $G^{\rm TX}_{n}$, $n=1,\ldots,\Nb$, besides considering the RSS or the SNR, we also need to understand the impact of interference when computing the coverage. To this aim, we will compute  the interference as the RSS of the signals transmitted through undesired beam-patterns
\begin{equation}
I[dW] = P_{\rm TX}[dBw]-LP_{\rm cable}[dB]+G_{\rm TX}[dB]-LP_{\rm at}[dB]-LP_{\rm fs}[dB]+G_{\rm RX}^{\rm interf}[dB]
\end{equation}
with $G^{\rm TX}_{n_{\rm selected}}$ the gain corresponding to the desired beampattern and  $G_{\rm RX}^{\rm interf} = \sum_{n\neq n_{\rm selected}}G^{\rm TX}_{n}$. With these definitions, the signal to interference plus noise ratio (SINR) has the expression
\begin{equation}
SINR = \frac{RSS[W]}{I[W]+\sigma^2[W]}.
\end{equation}

\subsection{Throughput and Spectral efficiency}
We consider the throughput expression given by the capacity $B_{\rm DL}\log_2(1+SINR)$ in bits per second.
We also consider the throughput without interference $B_{\rm DL}\log_2(1+SNR)$ as a bound to what can be achieved with inter-beam interference mitigation techniques. Assuming that frequency division duplexing mechanisms enable beam sharing among close users illuminated by the same beam, we will also consider the spectral efficiency with interference $\log_2(1+SINR)$ and without interference $\log_2(1+SNR)$ as performance metric. 

\section{Simulation Results}\label{sec:simulations}
Unless otherwise specified, we consider a LEO satellite orbiting at a $1300$ km height, covering an ellipsoidal  ROI with semi-radiuses $534.1$ km and $170.5$ km. This area has been designed following the procedure described in 3.1 for a LEO constellation with  83 orbits with an inclination of $53^\circ$ and 53 satellites in each orbit.
The downlink operates in the Ku band with a carrier frequency $f_\text{DL} = 11.45 {\rm GHz}$ and a bandwidth $B_\text{DL}= 250 {\rm MHz}$.
A single polarization is assumed in the simulations. 

\begin{table}[h!]
\begin{center}
\begin{tabular}{ |l|c|c|c| }
 \hline \hline
 Parameter & Symbol & Value & Units \\ \hline \hline 
 \multicolumn{4}{|c|}{\em Constellation and ROI} \\ \hline 
 Satellite height  & $h_{\rm sat}$ & $1300$ & km \\
  Number of orbital planes & $N_{\rm p}$ & $83$  & \\
  Number of satellites per orbital plane & $N_{\rm s}$ & $53$ & \\
  Orbital plane inclination & $\theta_\text{op}$ & $53$ & degrees \\ 
  Semiradius of the ROI in $x$-dimension  &  & 534.1 & km \\
  Semiradius of the ROI in $y$-dimension  &  & 170.5 & km \\  \hline 
   \multicolumn{4}{|c|}{\em Channel} \\ \hline  
Carrier frequency & $f_{\rm DL}$ & $11.45$ & GHz \\
 Bandwidth & $B_{\rm DL}$ & $250$ & MHz \\
  Atmospheric path loss & $LP_{\rm at}$ & $0.017$ & dB \\
 Rician factor  & $K_\text{r}$ & $10$ &  \\ \hline
 \multicolumn{4}{|c|}{\em Satellite}  \\ \hline
 Number of array elements  in $x$-dimension & $N^{\rm sat}_{\rm x}$ & $60$ & \\
  Number of array elements  in $y$-dimension  & $N^{\rm sat}_{\rm y}$ & $72$ & \\
  Number of RF chains in $x$-dimension & $N^{\rm RF}_{\rm x}$ & $5$  & \\
   Number of RF chains in $y$-dimension & $N^{\rm RF}_{\rm y}$ & $3$ & \\
 Transmit power & $P_{\rm TX}$ & $15$ & dBW \\
Oversampling factor  & $O$ & $1.2$ \\ \hline
 \multicolumn{4}{|c|}{\em User terminal } \\ \hline
Number of array elements  in $x$-dimension for UPA/Leaky wave antenna & $N^{\rm UT}_{\rm x}$ & $24$ & \\
Number of array elements  in $y$-dimension  for UPA/Leaky wave antenna & $N^{\rm UT}_{\rm y}$ & $24$ & \\
Kymeta antena peak gain & $G_\text{RX}$ & 33 & dB\\
Kymeta antena roll of factor & & 1.2 &   \\ 
Receiver noise temperature & T & 24.1 & dBK\\ \hline
\end{tabular}
\caption{System parameters for the simulation of the downlink of a LEO SatCom system operating in the Ku band.}\label{tab:system_param}
\end{center}
\end{table}

\begin{figure}
\begin{center}
    \begin{tabular}{cc}
     \includegraphics[width = 0.4\linewidth]{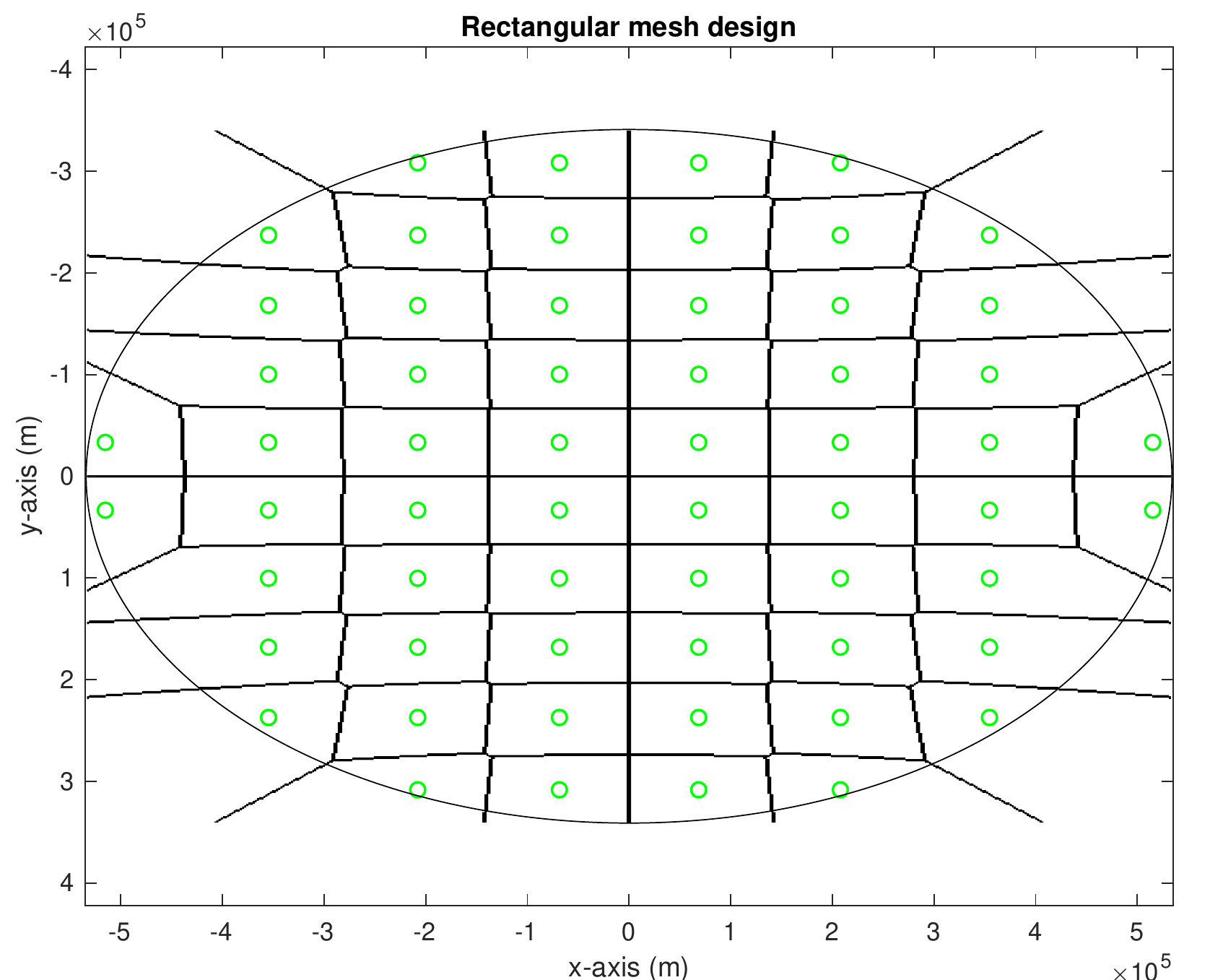} &
    \includegraphics[width = 0.4\linewidth]{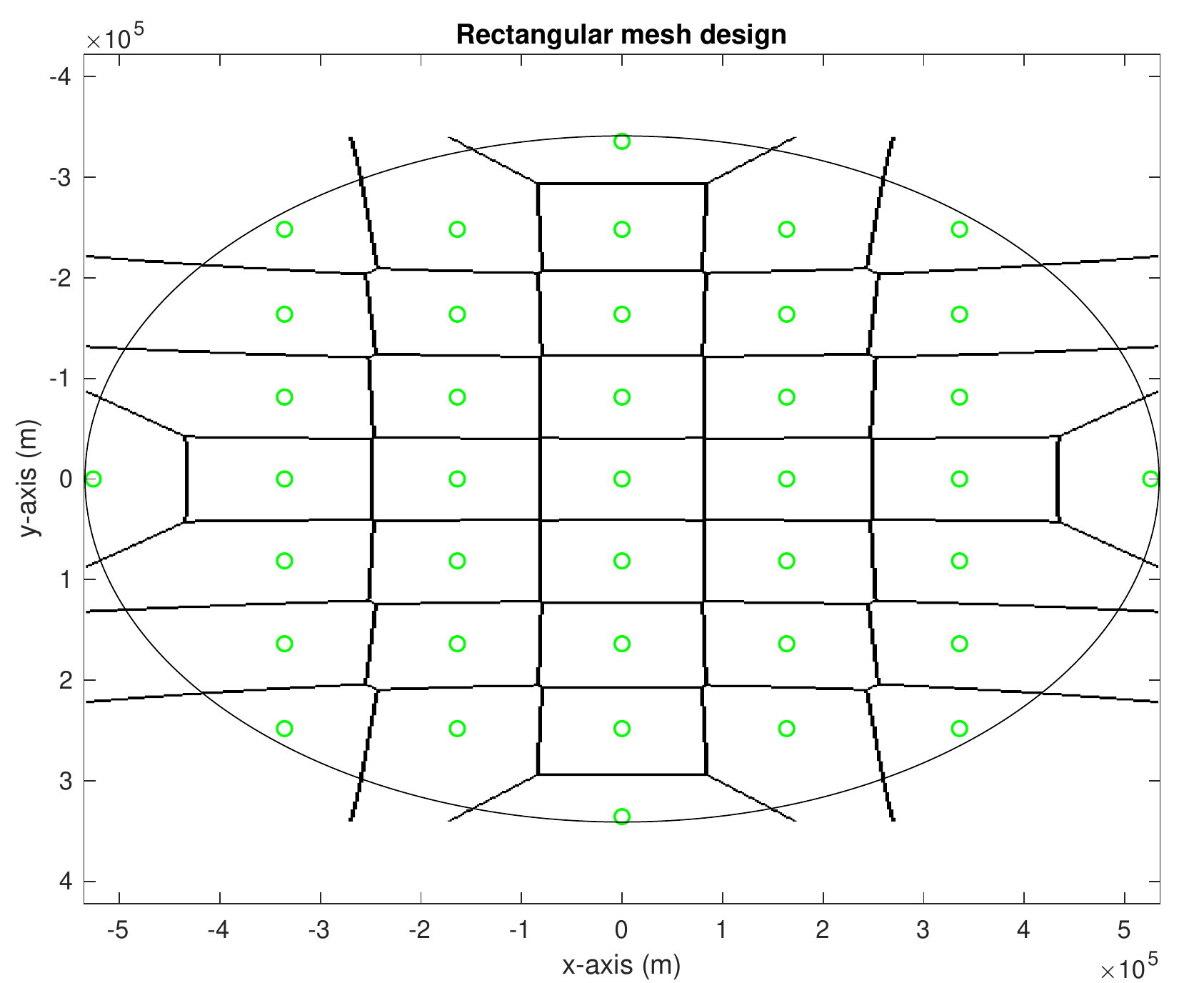} \\
    (a) & (b) \\
    \end{tabular}
    \caption{Rectangular mesh covering the elliptical ROI when using the oversampled DFT beam codebook: (a) with  $O=2.49$; (b) with $O=2$.  }
    \label{fig:mesh_vs_O}
\end{center}
\end{figure}

To understand the role of the oversampling factor $O$ in the  beam codebook presented in Section~\ref{sec:codebook}, we show in Fig.~\ref{fig:mesh_vs_O} different mesh designs covering the elliptical ROI for a  $128 \times 64$ satellite phased array and $8$ RF chains in each dimension. Fig.~\ref{fig:mesh_vs_O}(a) shows the results when $O=2.49$, while Fig.~\ref{fig:mesh_vs_O}(b) shows the obtained rectangular mesh when $O=2$. Note that with $O=2$, only 39 out of 64 potential beams are used to illuminate the ROI, resulting in a very inefficient usage of the RF resources.

For the next simulations we will assume that the satellite phased arrays consist of $5\times3$ sub-arrays of size $12\times24$, with a total of $60\times72$ antennas. Note that while we are arranging the sub-arrays into a larger plannar antenna array structure,  each subarray will be transmitting an independent signal. 
We found out that,  for this number of antennas and RF-chains, with an oversampling factor  of $1.2$ we can adjust all the available beams to the size of the ROI.
The remaining system parameters are summarized in Table~\ref{tab:system_param}.

\begin{figure}
    \begin{tabular}{ccc}
     \includegraphics[width = 0.32\linewidth, trim={4cm 7.5cm 4cm 6.5cm}, clip]{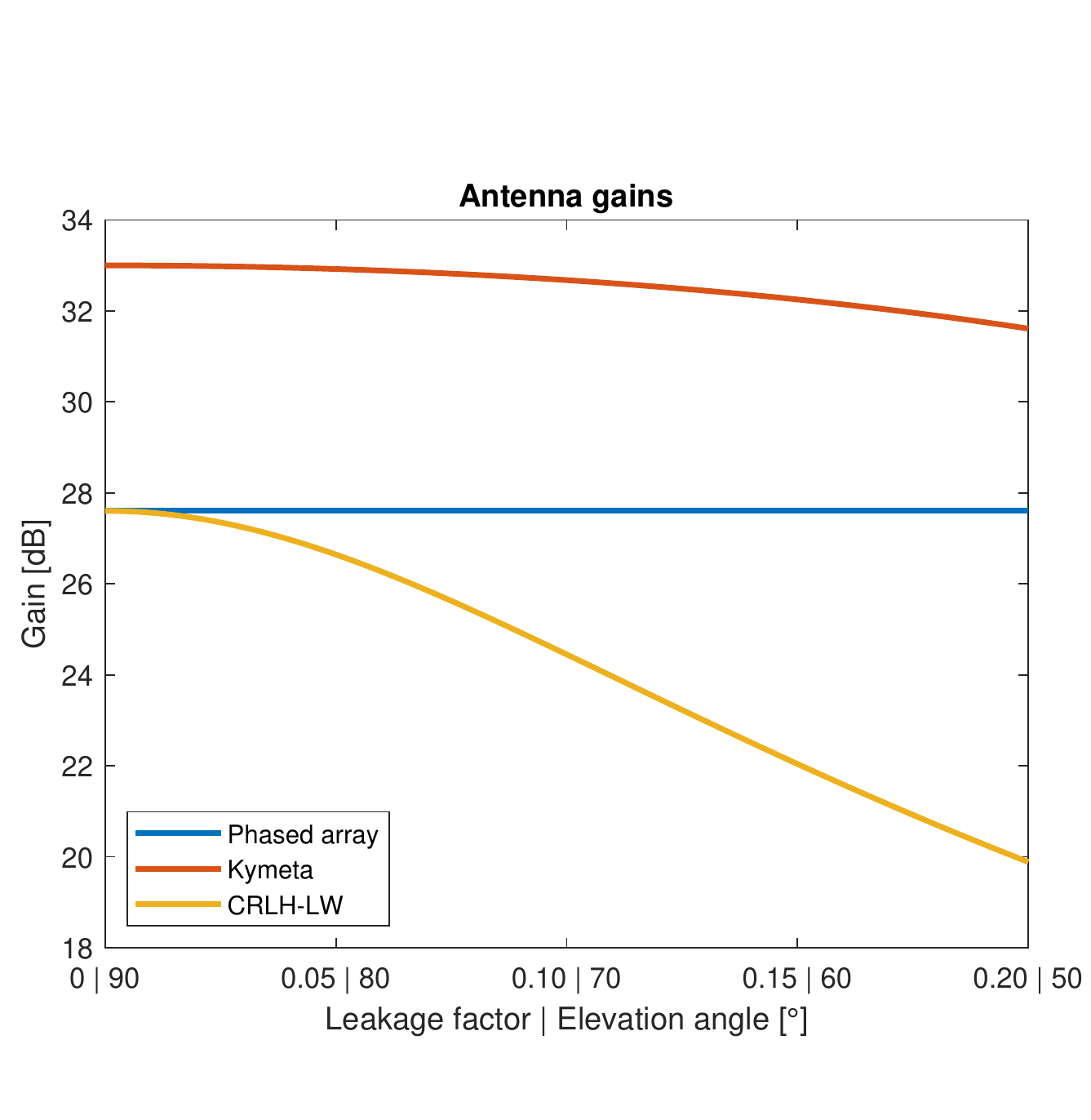} &
    \includegraphics[width = 0.32\linewidth, trim={4cm 7.5cm 4cm 6.5cm}, clip]{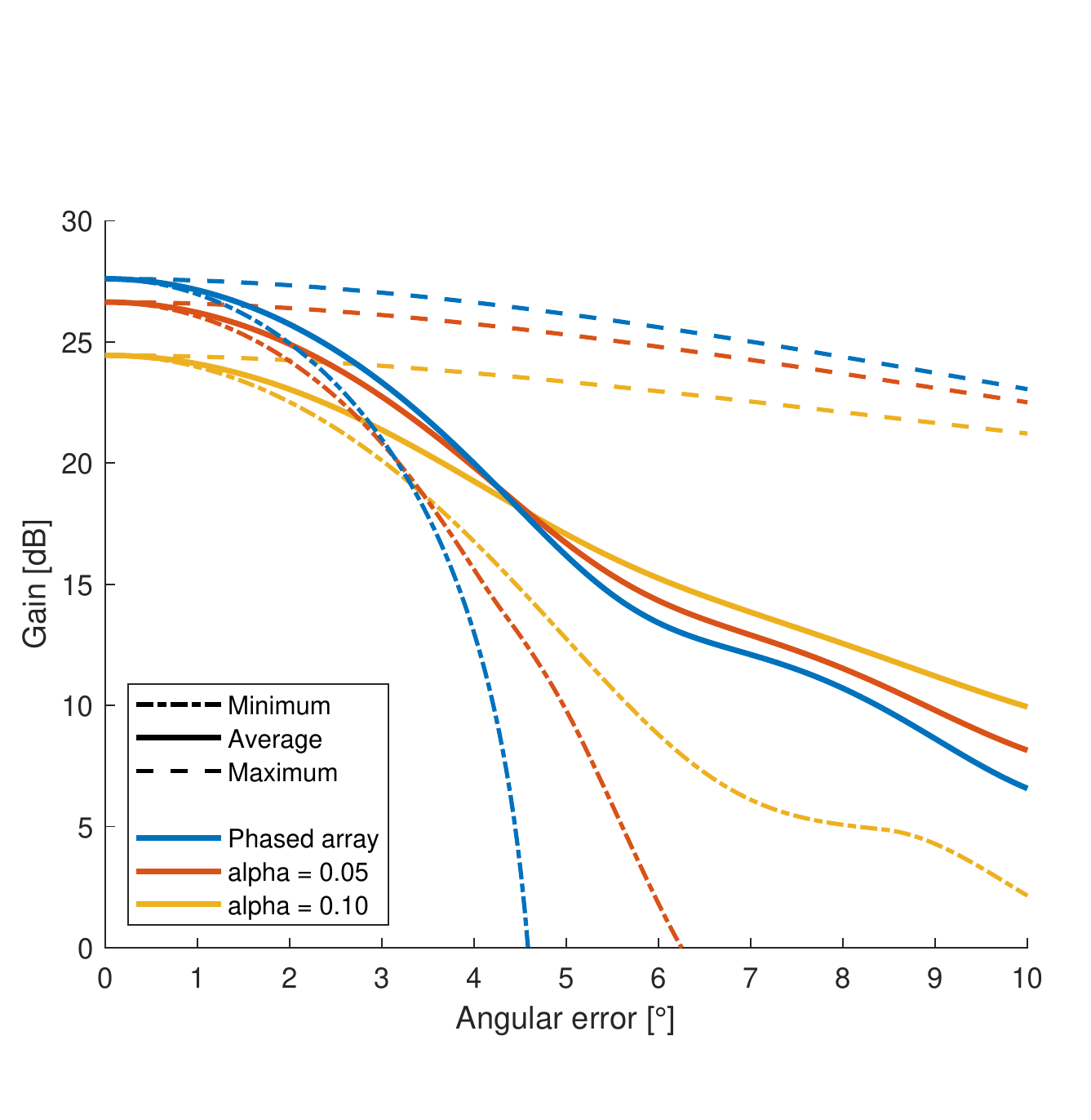} &
    \includegraphics[width = 0.32\linewidth, trim={4cm 7.5cm 4cm 6.5cm}, clip]{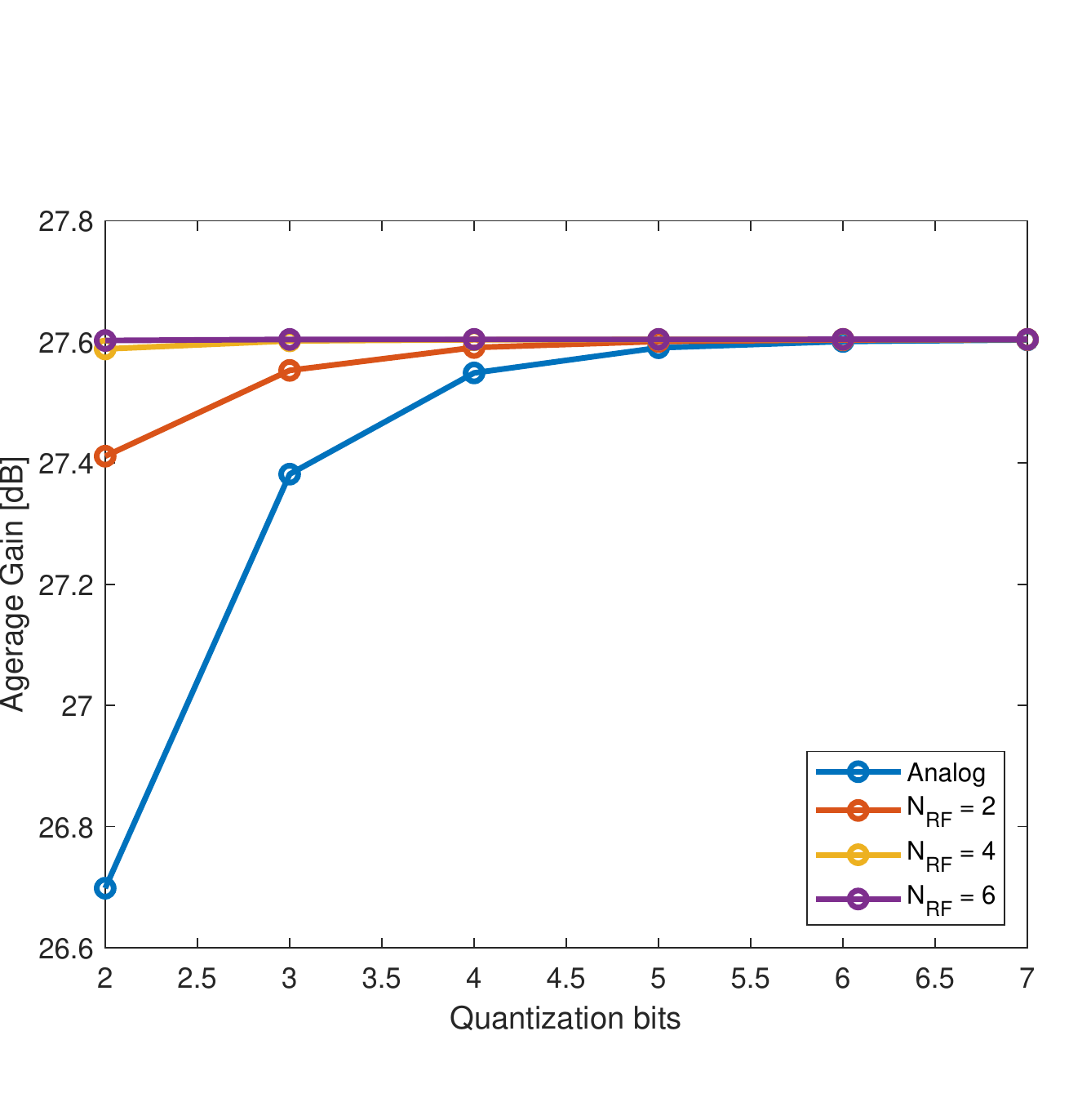} \\
    (a) & (b) & (c) \\
    \end{tabular}
    \caption{Gain provided by the different antenna terminals: (a) Gain with perfect beam alignment. (b) Gain under miss-alignment. (c) Gain under phase quantization.}
    \label{fig:user_gain}
\end{figure}

To understand the impact of different antennas at the user terminal, we will consider both the UPA and the leaky wave antenna (denoted as CRLH-LW) to have a number of elements equal to $24\times24$. Beamforming with the planar antenna array at the user side introduces  a gain of $27.6$ dB. The gain provided by the leaky wave antenna is a function of the leakage factor, as illustrated in Fig.~\ref{fig:user_gain}(a). The dependance of the antenna developed by Kymeta with the elevation angle  \cite{Kymeta_link_budget} is also shown in Fig.~\ref{fig:user_gain}(a).  
We consider now the effect of beam miss-alignment on the gain through Monte Carlo simulation for the phased antenna array and the leaky wave antenna.  Fig. \ref{fig:user_gain}(b) shows that, on average, the gains stay within reasonable levels, but the minimum values become critical under an error greater than $4\deg$.
The effects of phase shifter quantization for the phased array  in a hybrid architecture using the beamforming method proposed in \cite{Palacios2016} are shown in  Fig.~\ref{fig:user_gain}(c). It can be seen that a hybrid architecture offers a much better performance than an analog one under phase quantization, but both perform very similarly when the phase shifters have a high resolution.

\begin{figure}
    \centering
     \begin{tabular}{ccc}
    \includegraphics[width = 0.3\linewidth, trim={4cm 7.5cm 4cm 6.5cm}, clip]{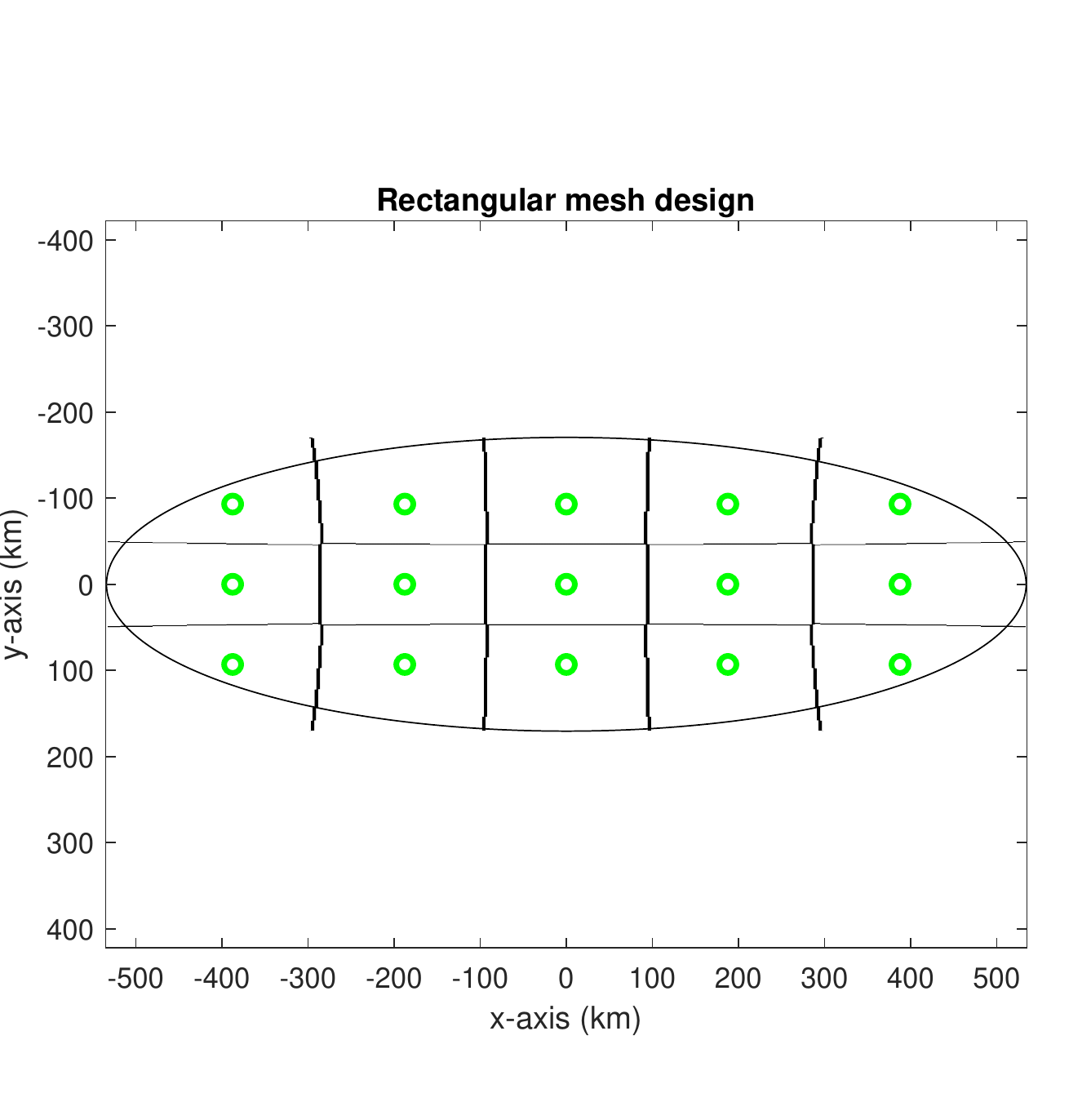}
    & \includegraphics[width = 0.3\linewidth, trim={3.8cm 7.5cm 4cm 6.5cm}, clip]{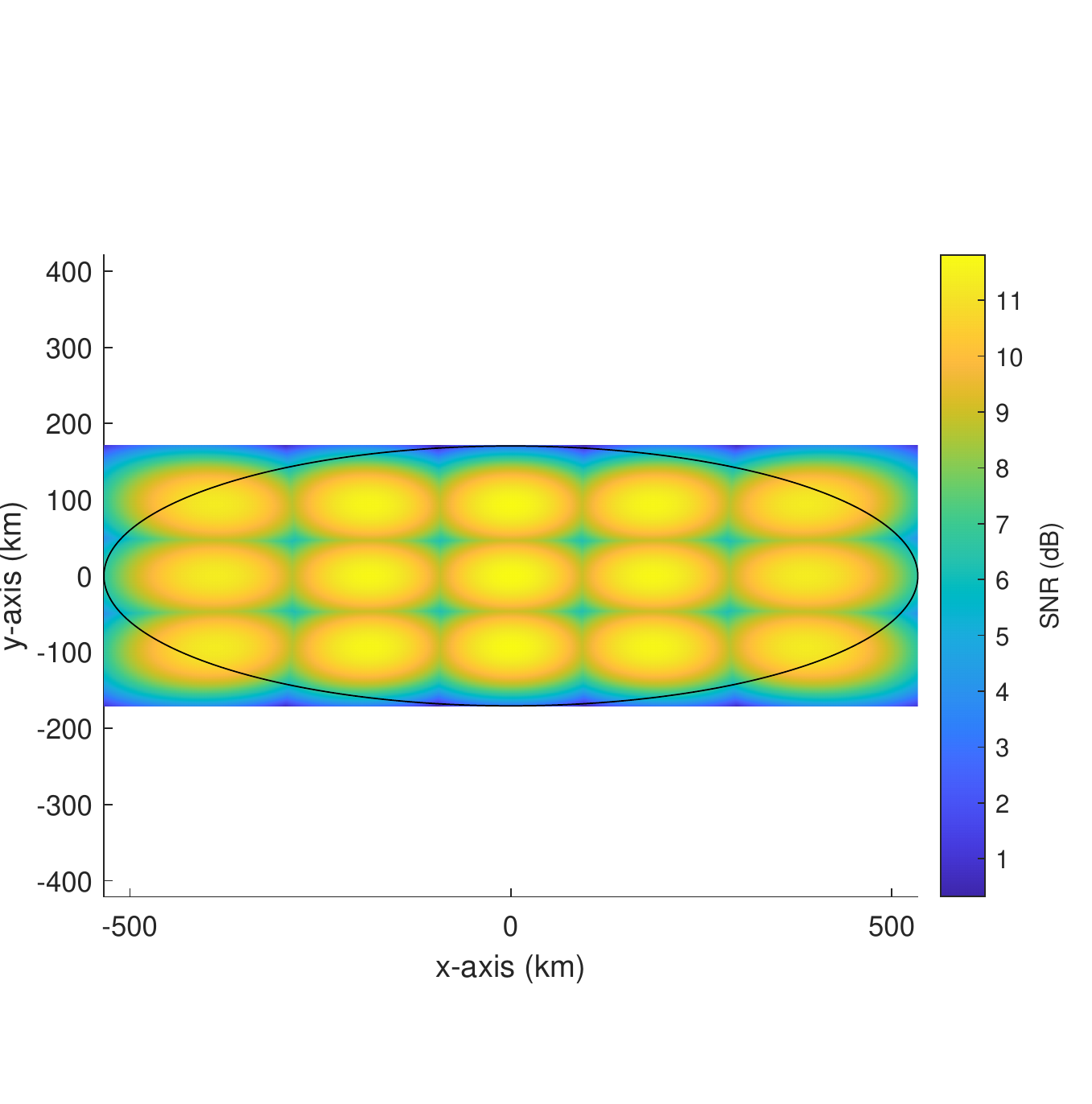}
    & \includegraphics[width = 0.3\linewidth, trim={3.8cm 7.5cm 4cm 6.5cm}, clip]{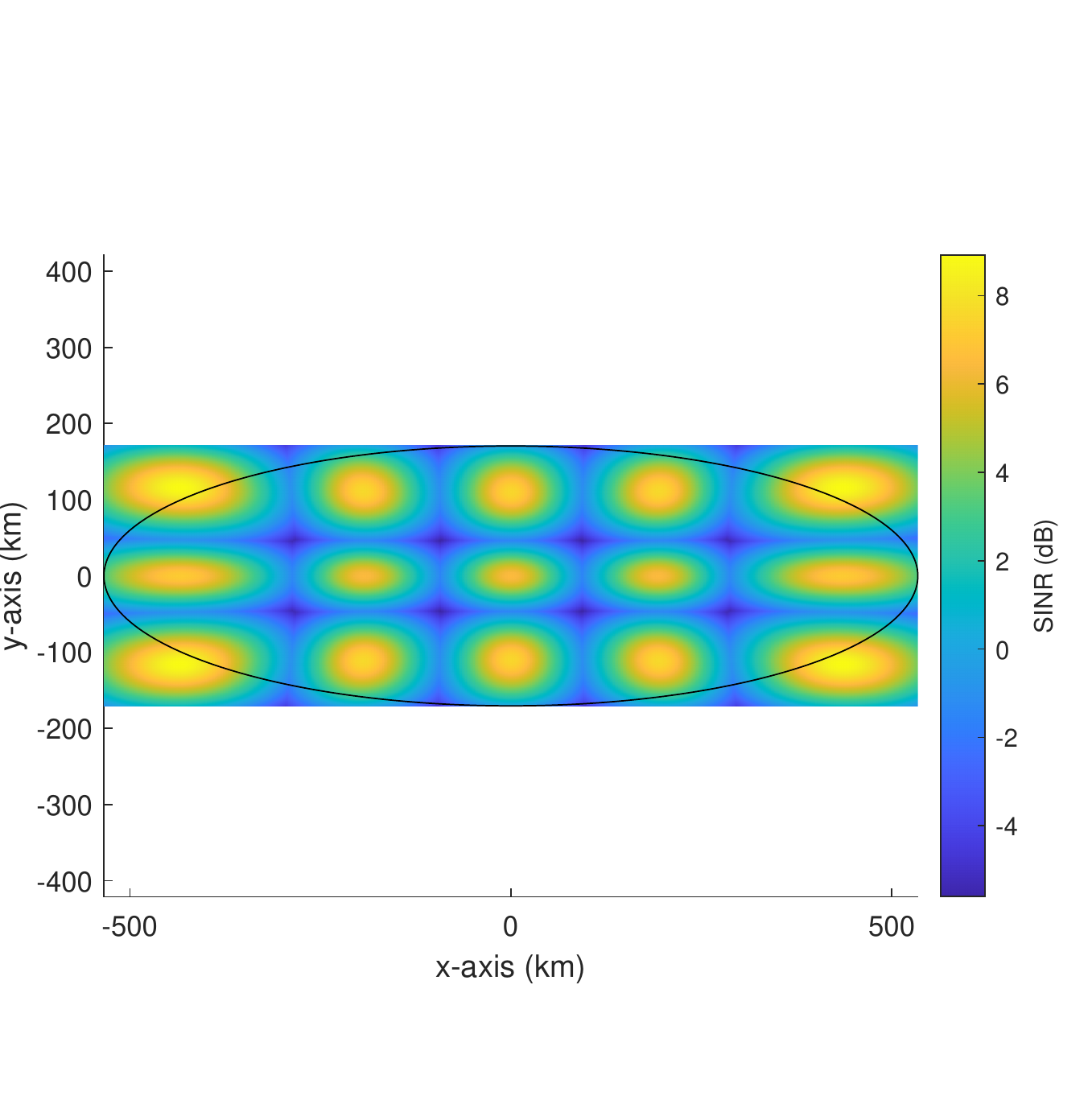}\\
    (a) & (b) & (c) \\
    \end{tabular}
    \caption{(a) Cells associated to the different beampatterns;  point marked in green show the location where the beampattern gain is maximum. (b) SNR when using the best beam for each cell. (c) SINR when using the best beam for each cell.}
    \label{fig:coverage}
\end{figure}
Next, we analyze the coverage of the elliptical ROI provided by the beam codebook at the satellite when the UPA is considered as antenna terminal.
As shown in Section~\ref{sec:coverage}, $LP_{\rm fs}$, $G_{\rm TX}$ and $G_{\rm TX}^{\rm interf}$ depend on the user location. The coverage provided by the satellite's codebook in terms of SNR and SINR can be seen in Fig.~\ref{fig:coverage}.
While the SNR shows a good coverage of the ROI for an oversampling factor of 1.2, when visualizing the SINR we can observe areas with high interference.
This is a consequence of  using the DFT-type codebook employed in 5G NR, since its associated rectangular mesh creates points that are covered with the same gain by 4 different satellite beams.  When selecting one of the beams, the other three will create a strong interference, leading to a SINR not larger than $-4.77{\rm [dB]}$. Furthermore, the design of static codebooks will always lead to points in which at least $3$ beam-patterns have equal gain, thus leading to a SINR never larger than $-3{\rm [dB]}$. 

\begin{figure}
    \centering
    \begin{tabular}{cc}
    \includegraphics[width = 0.4\linewidth, trim={3.8cm 7.5cm 4cm 6.5cm}, clip]{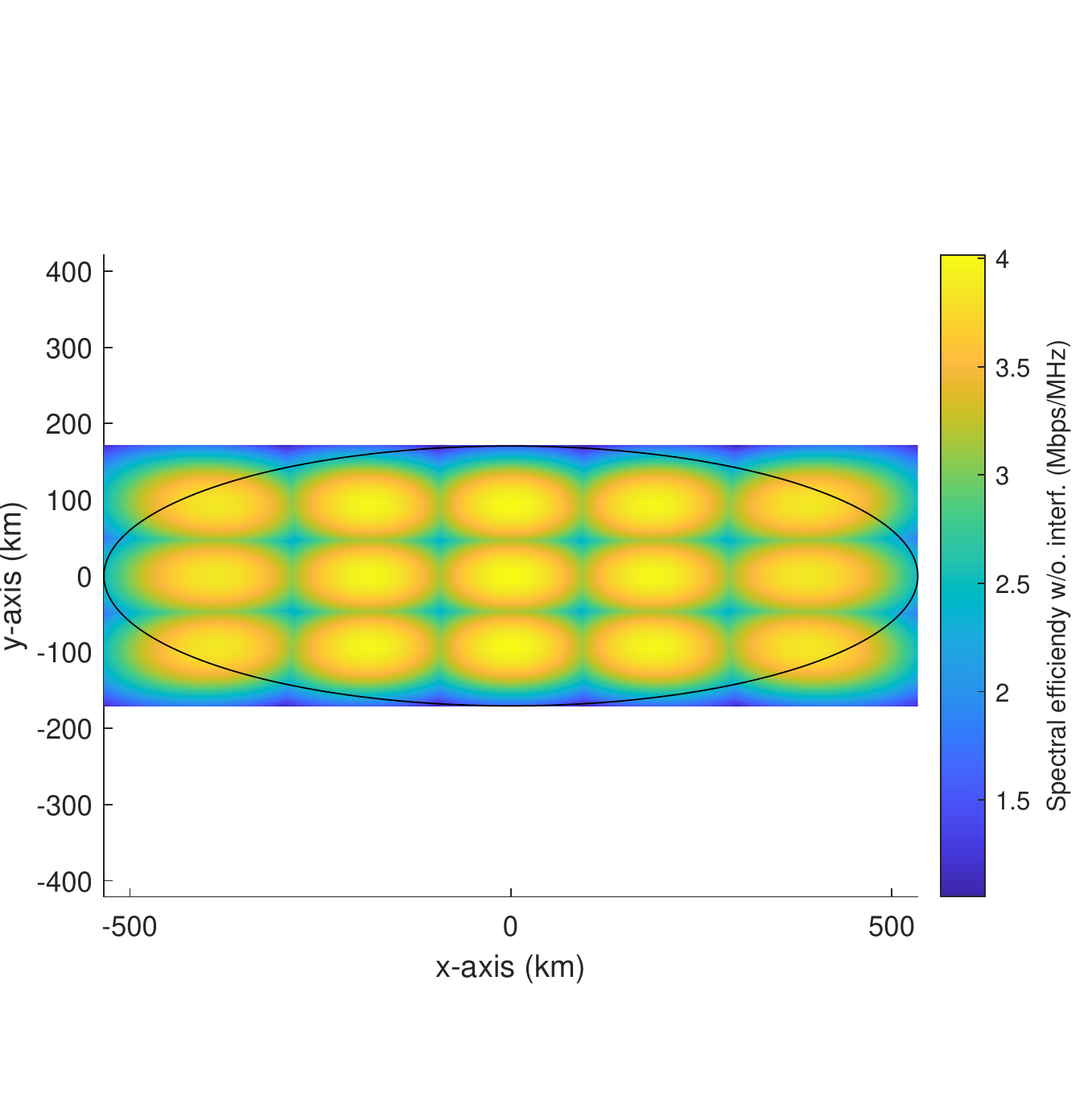} &
    \includegraphics[width = 0.4\linewidth, trim={3.8cm 7.5cm 4cm 6.5cm}, clip]{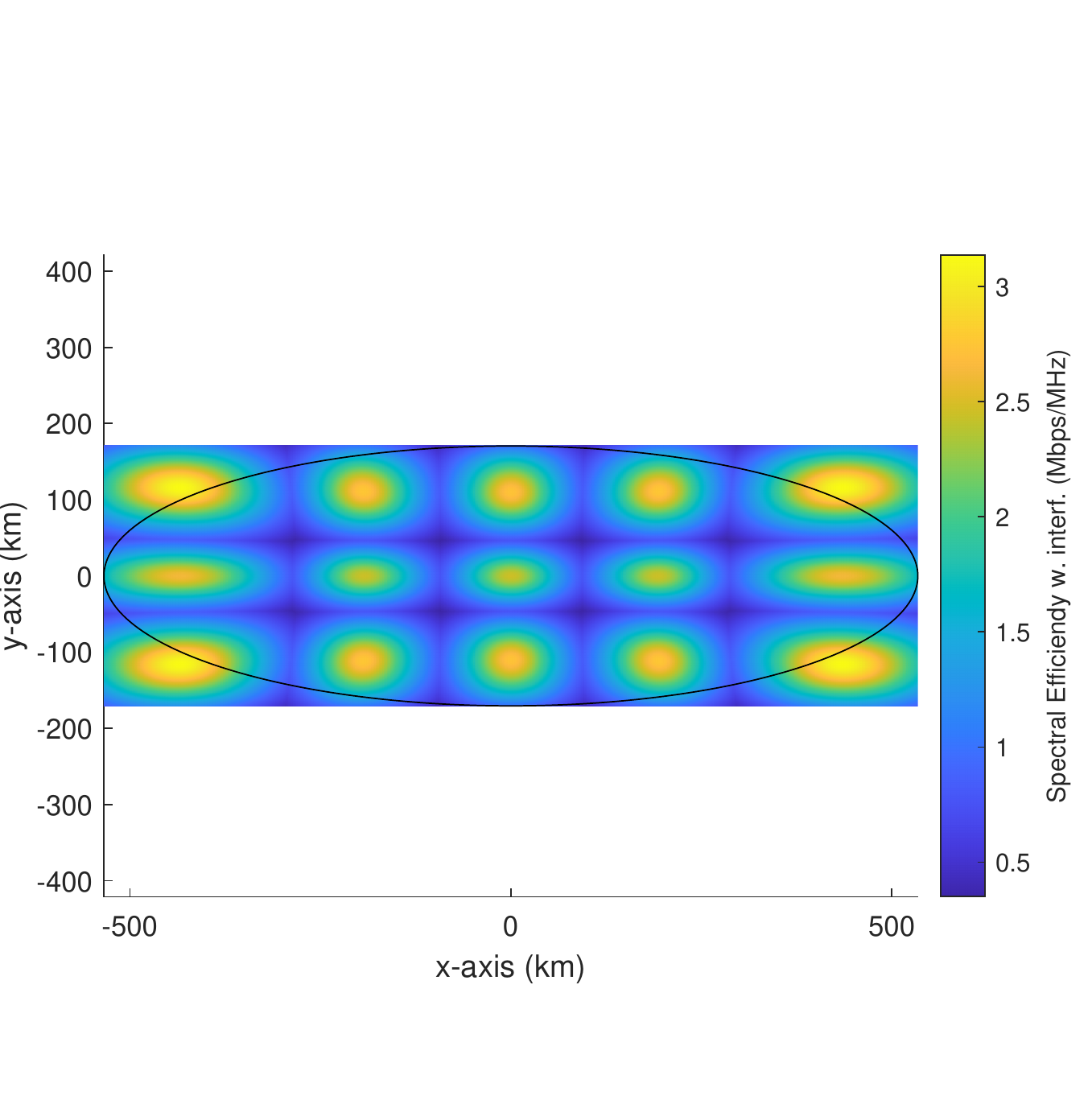}\\
    (a) & (b) \\
    \end{tabular}
    \caption{(a) Spectral efficiency when ignoring interference. (b) Spectral efficiency when considering interference.}
    \label{fig:throughput}
\end{figure}
\begin{figure}
    \centering
    \includegraphics[width = 0.4\linewidth, trim={4cm 7.5cm 4cm 6.5cm}, clip]{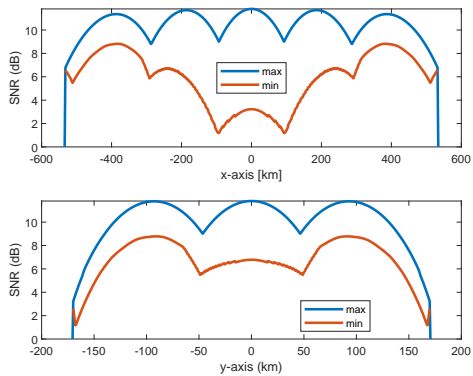}
    \caption{SNR range over both axis.}
    \label{fig:coverage_dim}
\end{figure}

In Fig. \ref{fig:coverage_dim} we see the SNR variation over each axis for the lines with maximum and minimum SNR. The curves with minimum SNR corresponds to the boundary of the elliptical ROI, while the curve for maximum SNR corresponds to the line $y=0$ in the coverage map.  It can be observed that the gain difference between the center of a cell and the edge is in the order of 3dB for both axes if the boundary of the ROI is excluded.
This SNR and SINR translate into the spectral efficiency shown in Fig.~\ref{fig:throughput}. Note that there is a significant inter-beam interference when operating with DFT-type codebooks as proposed in LTE and 5G NR. Future work needs to be developed to design alternative analog codebooks and hybrid beamforming strategies at the satellite side to reduce inter-beam interference.

\begin{figure}
    \centering
    \includegraphics[width = 0.3\linewidth, trim={4cm 7.5cm 4cm 8.5cm}, clip]{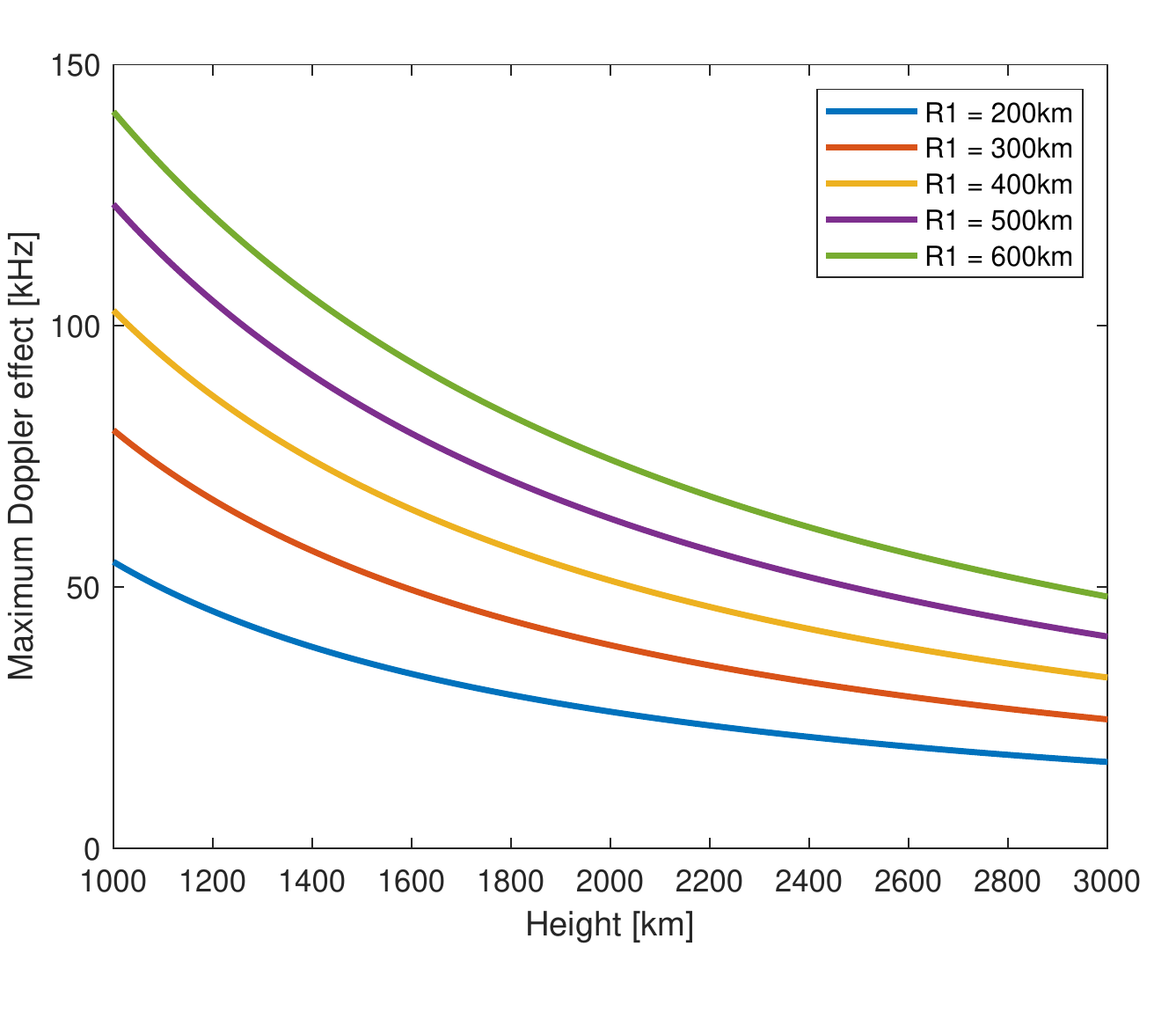}
    \includegraphics[width = 0.3\linewidth, trim={4cm 7.5cm 4cm 8.5cm}, clip]{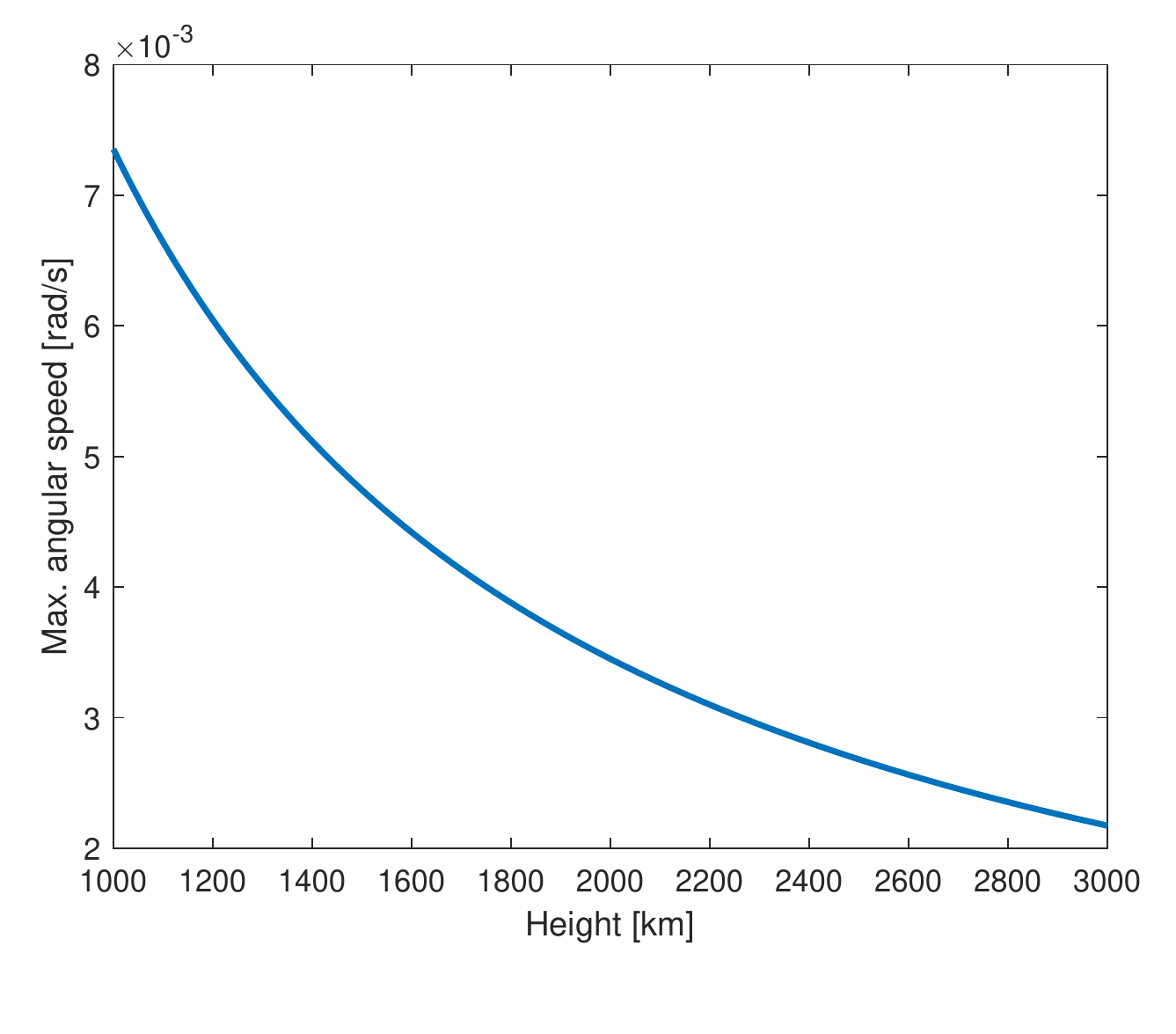}
    \includegraphics[width = 0.3\linewidth, trim={4cm 7.5cm 4cm 8.5cm}, clip]{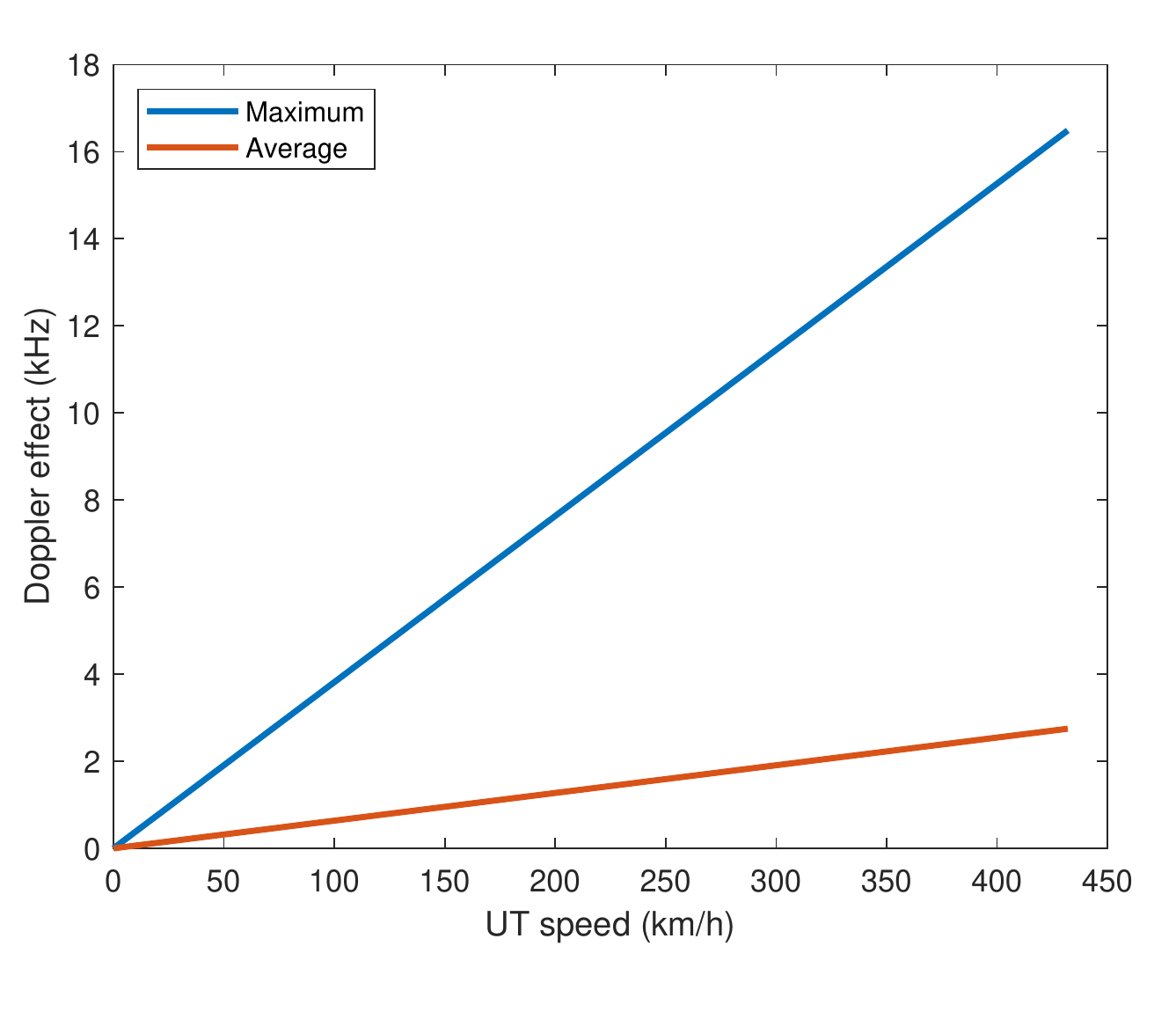}\\
    \vspace{-0.4cm}
    (a) \hspace{4.6cm} (b) \hspace{4.6cm} (c) \hspace{0.5cm}
    \caption{(a) Doppler effect caused by the satellite's movement for the carrier frequency. (b) Satellite's angular relative speed with respect to the user terminal. (c) Doppler effect caused by the UT's movement for the central frequency.}
    \label{fig:doppler-aspeed-theo}
\end{figure}

Fig. \ref{fig:doppler-aspeed-theo} shows the satellite's angular speed relative to the user and the Doppler effect due to the satellite's movement as a function of the user's relative location. 
The relative angular speed shows significantly small values. This means that the beampattern switching requirements of a given antenna design that enable tracking of the satellite's position for beamforming are not particularly challenging. Note that this holds when tracking the satellite direction, not its corresponding azimuth and elevation angles.  If the angular relative speed is translated into a maximum azimuth and elevation angular variation with time, the azimuth angle can present sharp changes, as illustrated in Section.~\ref{sec:Doppler_angular}. 

\begin{figure}
    \centering
    \begin{tabular}{cc}
     \includegraphics[width = 0.3\linewidth]{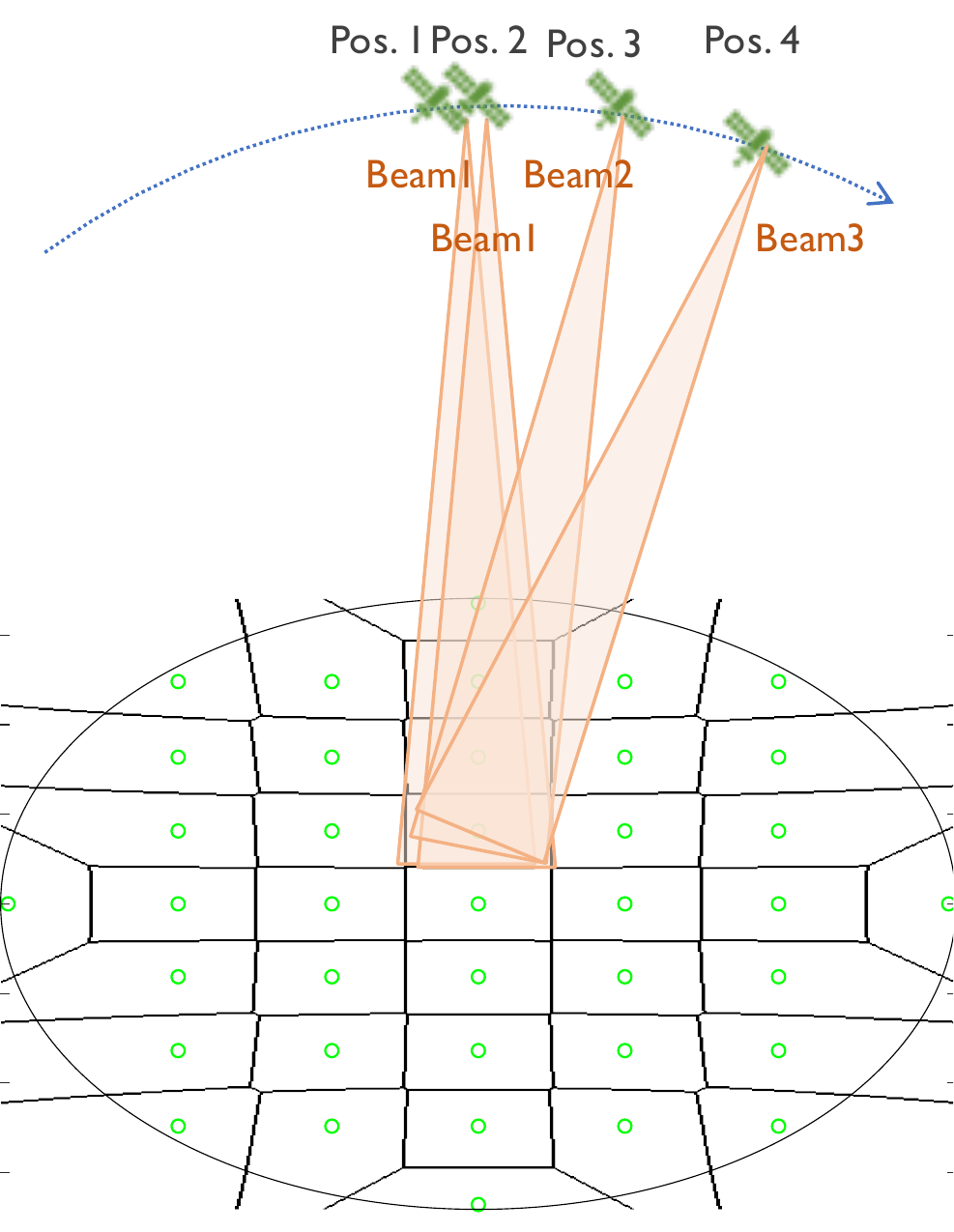} 
     & \includegraphics[width = 0.4\linewidth]{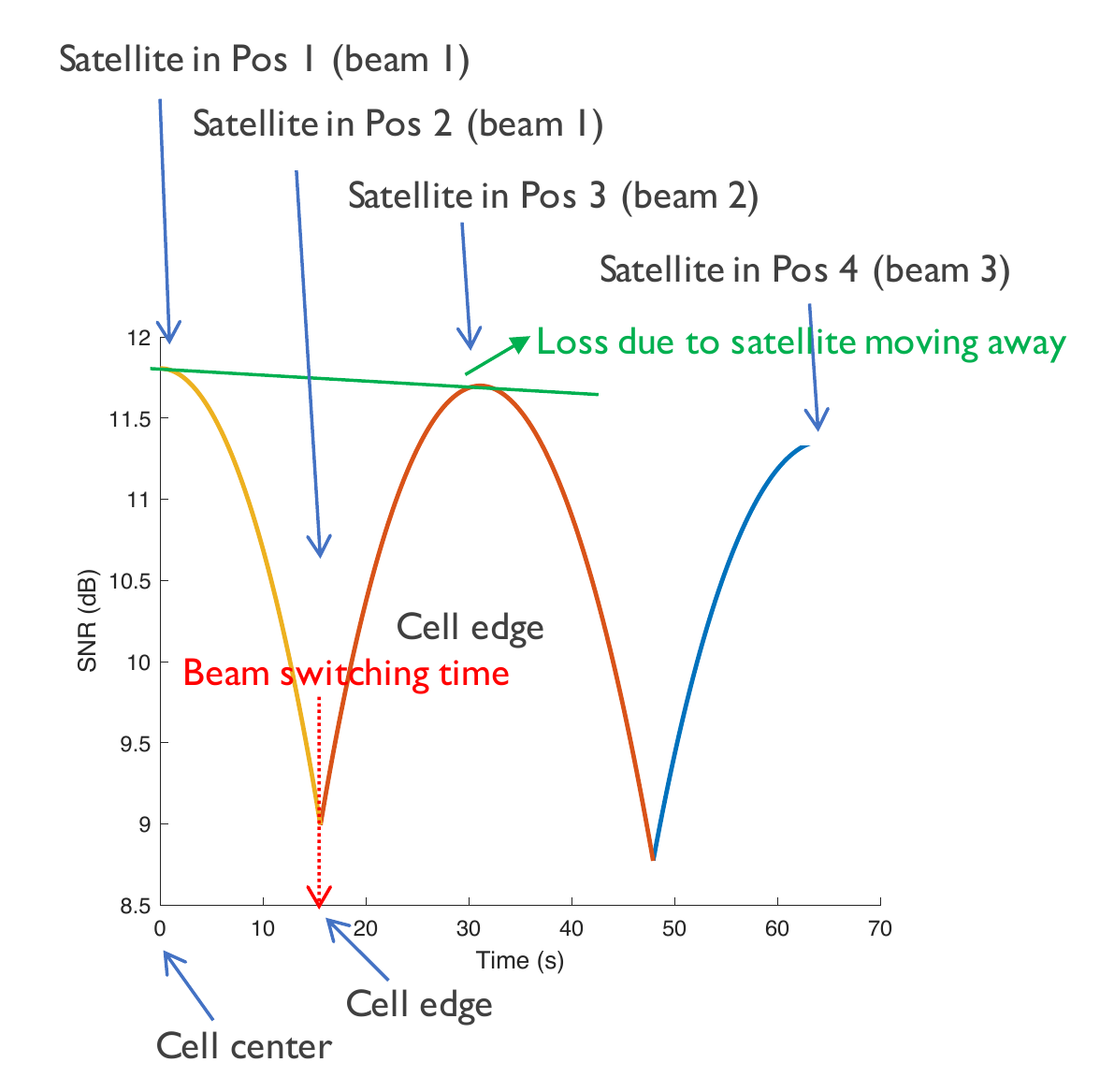}\\
    (a) & (b) \\
    \end{tabular}
    \caption{(a) Example of different satellite positions and associated beams. (b) SNR degradation over time due to the satellite's movement.}
    \label{fig:degrade}
\end{figure}

Finally, we analyze the SNR degradation over time due to the satellite's movement for a user initially located at the center of the elliptical ROI.
We have considered a beam selection strategy based on maximizing SNR.
Fig. \ref{fig:degrade}(a) shows an example of different satellite positions and associated selected beams. Fig. \ref{fig:degrade}(b) shows the SNR corresponding to the different satellite positions illustrated in (a).  It can be concluded that, for this codebook design and system parameters, the user is illuminated by the same  beam for a period of around $30$ seconds. After that, another beam has to be selected to keep a proper gain. Note that even after updating the best beam, there is a non-negligible gain loss due to the satellite movement, as also illustrated in Fig. \ref{fig:degrade}(b). 

\section{Conclusion}
Research on integrating LEO-based satellite networks with terrestrial networks is still in its infancy. We implemented a complete simulator of the downlink of a massive MIMO LEO SatCom system to understand the physical layer challenges and evaluate a hybrid beamforming strategy based on an analog codebook. We designed the satellite footprint as an elliptical ROI given the number of orbital planes and number of satellites per orbit. We also designed a static DFT-type codebook, as commonly used in current cellular standards, to adjust the beamwidth and RF resources to the size of the satellite footprint by means of an interpolation factor. Simulation results show the relevance of the antenna terminal design to achieve an acceptable SNR. Our numerical experiments also show that DFT codebooks introduce a high inter-beam interference which reduces the ability of the system to provide a high spectral efficiency at all locations inside the ROI. Designing fully or partially connected hybrid beamforming architectures with per antenna power constraints and a digital precoder that mitigates inter-beam interference is an open research challenge that needs to be addressed to overcome the SINR performance of current approaches. The design of low complexity dynamic codebooks that better adapt the resources to the specific location of the users and reduce the SNR loss due to the satellite movement is a critical challenge.  Our simulator also shows that the variation of the relative angular speed between the satellite and the UT is smooth when working with spherical coordinates. This  leads to the need of designing beam tracking methods for the UT that track the satellite position rather than its variation in elevation and azimuth. Our numerical experiments also show the effectiveness of exploiting satellite position information at the UT side to avoid the need of channel estimation and tracking, which will provide old estimates due to the long transmission delay. Finally, an accurate model for the error in the satellite's position and trajectory is also necessary to better understand its impact in the performance of the satellite beam tracking strategies.
\bibliographystyle{IEEEtran}


\end{document}